\documentstyle[lprocl,epsfig,epsf,wrapfig,rotating,11pt]{article}

\def\be{\begin{equation}}
\def\ee{\end{equation}}
\def\bea{\begin{eqnarray}}
\def\eea{\end{eqnarray}}
\def\be{\begin{equation}}
\def\ee{\end{equation}}
\def\bea{\begin{eqnarray}}
\def\eea{\end{eqnarray}}
\def\ee{\mbox{e}^+\mbox{e}^-}

\def\sqee{\sqrt{s}_{\rm ee}}

\def\gg{\gamma\gamma}

\def\ee{\mbox{e}^+\mbox{e}^-}

\def\ETJET{E^{\rm jet}_T}
\def\ESJET{E^{*}_T}

\def\qqbar{\mbox{q}\overline{\mbox{q}}}
\def\ccbar{\mbox{c}\overline{\mbox{c}}}

\def\q{\mbox{q}}
\def\g{\mbox{g}}

\def\xgp{x_{\gamma}^+}
\def\xgm{x_{\gamma}^-}
\def\xgpm{x_{\gamma}^{\pm}}
\def\etajet{\eta^{\rm jet}}

\def\Zzero{\ifmmode {{\mathrm Z}^0} \else {${\mathrm Z}^0$} \fi}
\def\ppbar{\overline{\mbox p}\mbox{p}}

\def\sigmagg{\sigma_{\gg}}
\def\fl         {\ifmmode {F^{\gamma}_L}
                 \else    {$F^{\gamma}_L$}\fi}
\def\fb         {\ifmmode {F^{\gamma}_B}
                 \else   {$F^{\gamma}_B$}\fi}

\begin{document}

\pagestyle{myheadings}
\markboth{FREIBURG-EHEP-97-19}{FREIBURG-EHEP-97-19}

\setcounter{footnote}{0}
\renewcommand{\thefootnote}{\fnsymbol{footnote}}
\title{THE STRUCTURE OF THE PHOTON\footnotemark[1] }

\author{STEFAN S\"OLDNER-REMBOLD}

\address{Albert-Ludwigs-Universit\"at Freiburg, Hermann-Herder-Str.~3\\
D-79104 Freiburg, Germany \\E-mail: soldner@ruhpb.physik.uni-freiburg.de}


\maketitle\abstracts{The structure of the photon is studied
in high energy photon-proton interactions at HERA and
photon-photon interactions at LEP. The status of these
measurements is reviewed.}

\section{Introduction}
The photon is one of the fundamental gauge bosons of the Standard Model
without self-couplings and without intrinsic structure. 
However, at high energies photon-hadron and photon-photon interactions are 
dominated by quantum fluctuations
of the photons into fermion-antifermion pairs and into
vector mesons which have the same spin-parity ($J^{PC}=1^{--}$)
as the photon. This is called photon structure. 
Electron-positron collisions at LEP and positron-proton
collisions at HERA are an ideal laboratory
for studying photon structure in the interactions of quasi-real and virtual 
photons, 
testing predictions of both 
Quantum Electrodynamics (QED) and Quantum Chromodynamics (QCD). 
\setcounter{footnote}{0}
\renewcommand{\thefootnote}{\fnsymbol{footnote}}
\footnotetext[1]{Invited talk given at the XVIII International
Symposium on Lepton Photon Interactions,
Hamburg, Germany, July 28--August 1, 1997} 

\setcounter{footnote}{0}
\renewcommand{\thefootnote}{\alph{footnote}}
\section{Electron-photon scattering}
\begin{wrapfigure}[15]{l}{2.5in}
\epsfig{file=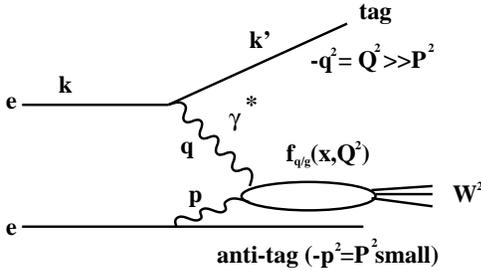,width=2.5in} 
\caption{\label{fig-egfig}
Deep-inelastic e$\gamma$ scattering:
$k(k')$ denotes the four-momentum of the incoming (scattered) electron
and $q(p)$ the four-momentum of the virtual (quasi-real) photon.}
\end{wrapfigure}

If one of the scattered electrons in $\ee$ collisions is detected (tagged), 
the process $\ee \rightarrow \ee + \mbox{hadrons}$ (Fig.~\ref{fig-egfig}) 
can be regarded as deep-inelastic scattering of an 
electron\footnote{In this paper positrons 
are also referred to as electrons}~on 
a quasi-real photon which has been radiated by the other electron
beam. The cross-section is written as
 \begin{equation}
  \frac{{\rm d}^2\sigma_{\rm e\gamma\rightarrow {\rm e+hadrons}}}{{\rm d}
x{\rm d}Q^2}
 =\frac{2\pi\alpha^2}{x\,Q^{4}}
  \left[ \left( 1+(1-y)^2\right) F_2^{\gamma}(x,Q^2) - y^{2}
F_{\rm L}^{\gamma}(x,Q^2)\right],
\label{eq-eq1}
 \end{equation}
where $\alpha$ is the fine structure constant and 
$$Q^2=-q^2=-(k-k')^2$$ 
is the negative four-momentum squared of the virtual photon $\gamma^*$ and 
$$x=\frac{Q^2}{2p\cdot q}=\frac{Q^2}{Q^2+W^2+P^2}$$ 
$$y=\frac{p\cdot q}{p\cdot k}$$
are the usual dimensionless variables of deep-inelastic scattering.
$W^2=(q+p)^2$ is the squared invariant mass of the hadronic final state.
The negative four-momentum squared, $P^2=-p^2$,
 of the quasi-real target photon is
approximately zero and therefore usually neglected.
In leading order (LO) the photon structure function $F_2^{\gamma}(x,Q^2)$ 
is related to the sum over the quark densities of the 
photon
weighted by the quark charge $e_{\rm q}$
$$F_2^{\gamma}(x,Q^2)=2x\sum_{\rm q} e^2_{\rm q} 
f_{\rm q/\gamma}(x,Q^2)$$ 
with $f_{\rm q/\gamma}(x,Q^2)$ being the probability to
find a quark flavour q with the momentum fraction $x$ 
(sometimes denoted by $x_{\gamma}$) in the photon. 
For measuring $F_2^{\gamma}(x,Q^2)$ the values 
of $Q^2$ and $y$ 
can be reconstructed from the energy, $E_{\rm tag}$,
and the angle, $\theta_{\rm tag}$, of the tagged electron and the beam
energy $E_{\rm beam}$ via the relations 
$$Q^2\approx 2E_{\rm beam} E_{\rm tag}(1-\cos\theta_{\rm tag})$$
$$y\approx 1-\frac{E_{\rm tag}}{E_{\rm beam}}\cos^2\frac{\theta_{\rm tag}}{2}.$$
In order to identify an electron in the detector, 
the tag energy $E_{\rm tag}$ has to be large, i.e.~only low values of 
$y$ are accessible ($y^2\ll 1$). 
The contribution of the cross-section term proportional to the 
longitudinal structure function $F_{\rm L}^{\gamma}$ is therefore negligible.
 
The reconstruction of $x$, however, relies heavily on the 
measurement of the invariant mass $W$ from the energies $E_{\rm h}$ and
momenta $\vec{p}_{\rm h}$ of the final state hadrons h:
$$W^2=\left(\sum_{\rm h} E_{\rm h}\right)^2-\left(\sum_{\rm h}
\vec{p}_{\rm h}\right)^2.$$
Unfolding of the $x$ dependence of the structure function therefore
requires that the hadronic final state in e$\gamma$ events
is well measured and well simulated by the Monte Carlo models.

\subsection{Hadronic energy flows}
OPAL~\cite{bib-hadopal}, ALEPH~\cite{bib-alephflow} and 
DELPHI~\cite{bib-igor} have studied 
the hadronic energy flow per event, $1/N\cdot {\rm d}E/{\rm d}\eta$,
as a function of the pseudorapidity $\eta=-\ln\tan\theta/2$, where
the sign of $\eta$ is chosen in such way that the
tag electron is always at negative $\eta$. OPAL has measured 
\begin{figure}[htbp]
\begin{tabular}{cc}
\epsfig{file=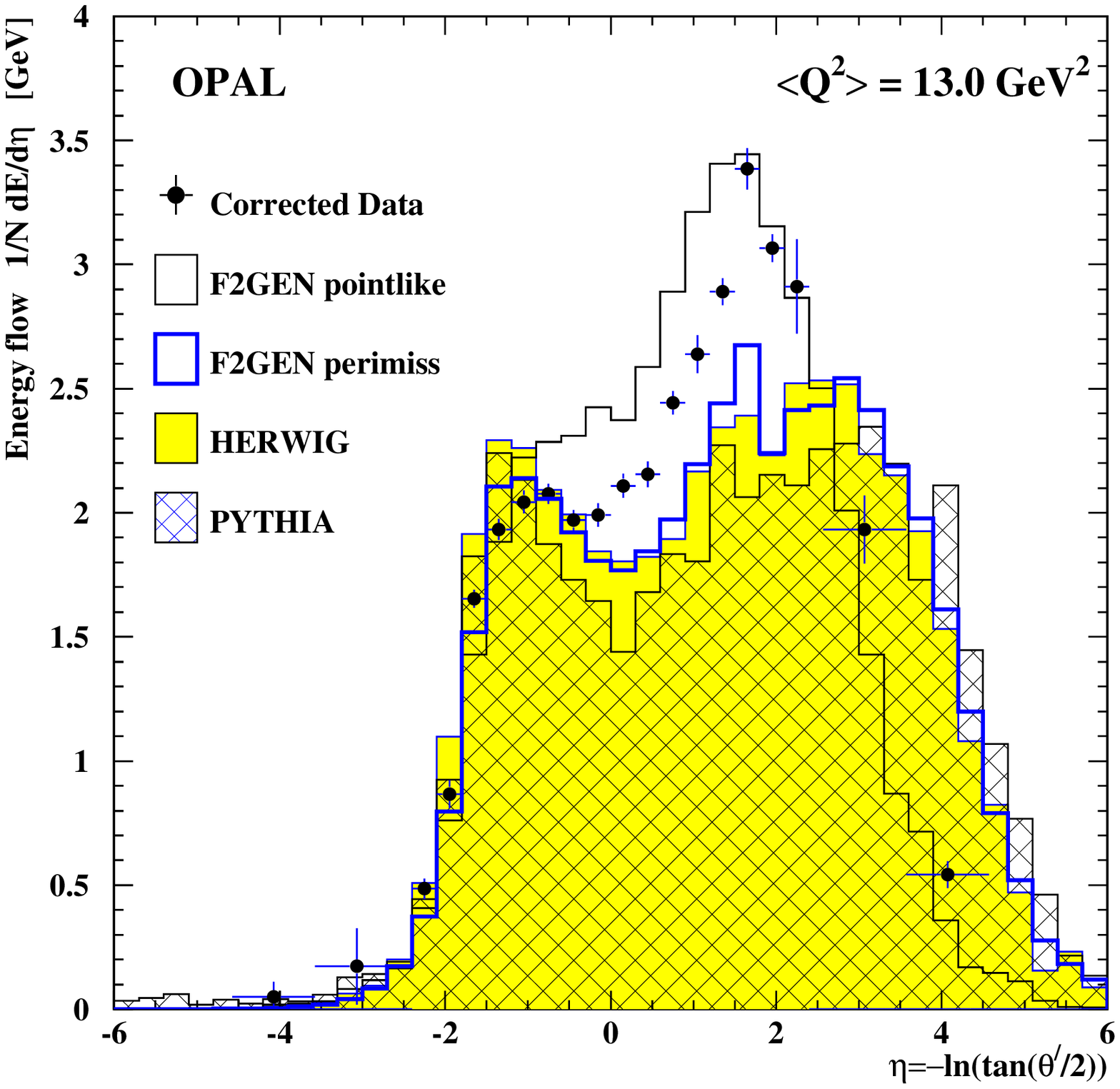,width=0.463\textwidth,height=5.3cm}
 &
\epsfig{file=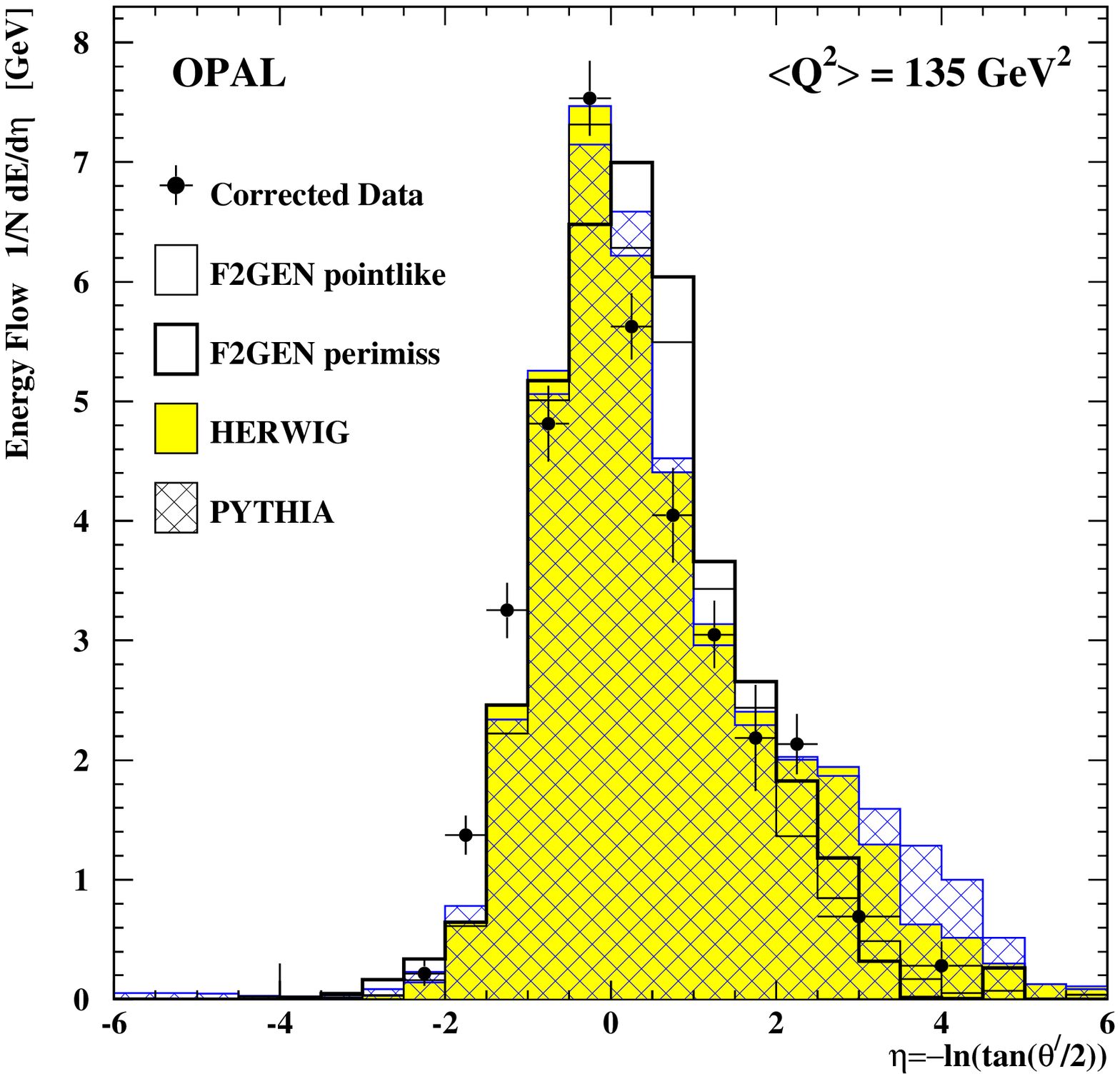,width=0.463\textwidth,height=5.3cm} 
\end{tabular}
\caption{\label{fig-hadopal}
 The energy flow per event
 as a function of the pseudorapidity $\eta$ in 
 different $Q^2$ ranges compared to
 the HERWIG, PYTHIA and F2GEN Monte Carlo models.
 The data have been corrected for detector effects.}
\end{figure}
the energy flow at medium $Q^2$ 
($\langle Q^2\rangle=13$ GeV$^2$) and at high $Q^2$
($\langle Q^2\rangle=135$ GeV$^2$)~\cite{bib-hadopal}. 
In Fig.~\ref{fig-hadopal} the energy flows are compared
to the two QCD based Monte Carlo generators HERWIG~\cite{bib-herwig}
and PYTHIA~\cite{bib-pythia}.
The data distributions have been corrected for detector effects.
The generator F2GEN is used
to model the unphysically extreme case 
of a two-quark state in the $\gamma^*\gamma$ centre-of-mass system
with an angular distribution as in lepton pair production from two real 
photons (``pointlike''). The ``perimiss'' sample is 
a physics motivated mixture of pointlike and peripheral
interactions, where peripheral means that the
transverse momentum of the outgoing quarks is given by
an exponential distribution as if 
all the photons interacted as pre-existing hadrons.

Significant discrepancies exist between the data and all of
the Monte Carlo models. The agreement improves at higher $Q^2$.
Since $x$ and $Q^2$ are correlated, the discrepancies
at low $Q^2$ are observed also at low $x$. These discrepancies
between the data and the Monte Carlo model for the hadronic final
state are the dominant source of systematic uncertainty
in the unfolding of $F_2^{\gamma}(x,Q^2)$~\cite{bib-hadopal}.

\begin{wrapfigure}{r}{2.5in}
\epsfig{file=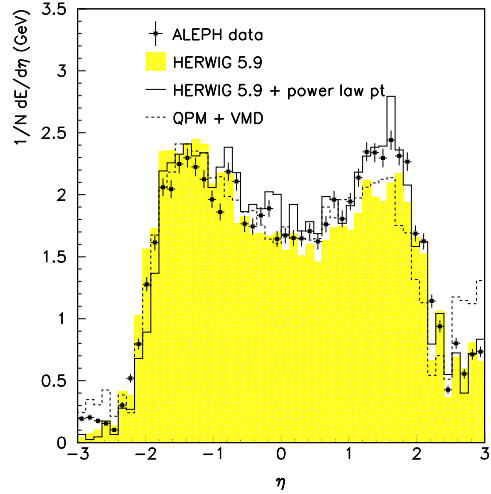,width=2.5in} 
\caption{\label{fig-alephflow}
 The energy flow per event
 as a function of $\eta$, compared to
 a QPM+VDM model,
 the standard and the tuned version of HERWIG.} 
\end{wrapfigure}
Tuning of the Monte Carlo to improve the modelling of
the hadronic final state is complicated and must be done with care
in order to avoid a bias of the result of the unfolding
towards the parametrisation of the parton distribution
functions used in the tuned Monte Carlo. ALEPH~\cite{bib-alephflow} has
measured the energy flow in tagged event for
$\langle Q^2 \rangle=14.2$~GeV$^2$.
The distributions have not been corrected for detector effects.
The energy flow shown in Fig.~\ref{fig-alephflow} is compared
to the HERWIG generator~\cite{bib-herwig} and a Monte Carlo
which consists of a mixture of Quark Parton Model (QPM)
and Vector Meson Dominance (VDM) similar to the F2GEN
generator. In addition, a modified version
of HERWIG (``HERWIG+power law $p_{\rm T}$'') was used.
The modification is based on studies of the photon remnant by 
ZEUS~\cite{bib-rzeus}. In standard HERWIG a Gaussian distribution is 
used to describe
the limitation of the transverse momentum of the outgoing partons
with respect to the initial target photon. In the modified
HERWIG the Gaussian is replaced by a power law spectrum.
The agreement with the data improves. A similar study
was performed earlier in Ref.~7 using OPAL data.
It is expected that such improvements of the Monte Carlo
models will significantly reduce the systematic error
of the structure function measurements for hadronic events.

\subsection{The photon structure function $F_2^{\gamma}$ at high $Q^2$}
\label{sec-egamma}
Even though the concept of the photon structure function 
$F_2^{\gamma}$ has
been developed in analogy to the formalism of the nucleon structure
functions $F_2^{\rm N}$, there are important differences: 
$F_2^{\gamma}(x,Q^2)$ increases with $Q^2$ for all $x$ and this
positive scaling violation is expected already within the parton model. 
Furthermore, $F_2^{\gamma}$ is large for high $x$, whereas
$F_2^{\rm N}$ decreases at large $x$. These differences are due
to the additional perturbative $\gamma\rightarrow\qqbar$ splitting
which does not exist for the nucleon. 

For large $x$ and asymptotically large $Q^2$ the value of
$F_2^{\gamma}$ can therefore be calculated from perturbative 
QCD~\cite{bib-witten}. The next-to-leading order (NLO) 
result~\cite{bib-buras} can be written as
\begin{equation}
\frac{F_2^{\gamma}}{\alpha}=\frac{a(x)}{\alpha_{\rm s}(Q^2)}+b(x),
\end{equation}
where $a(x)$ and $b(x)$ are calculable functions which diverge
for $x\rightarrow 0$ and $\alpha_{\rm s}$ is the strong coupling constant.
The first term corresponds to the LO
result by Witten~\cite{bib-witten}. The measurement
of $F_2^{\gamma}$ could be a direct measurement of $\Lambda_{\rm QCD}$
if it were not for the large non-perturbative contributions
due to bound states. These contributions are large at all experimentally
accessible values of $Q^2$.

\unitlength1cm
\begin{figure}[htbp]
\begin{picture}(15.0,6.5)
\put(0,3.25){\begin{tabular}{cc}
\epsfig{file=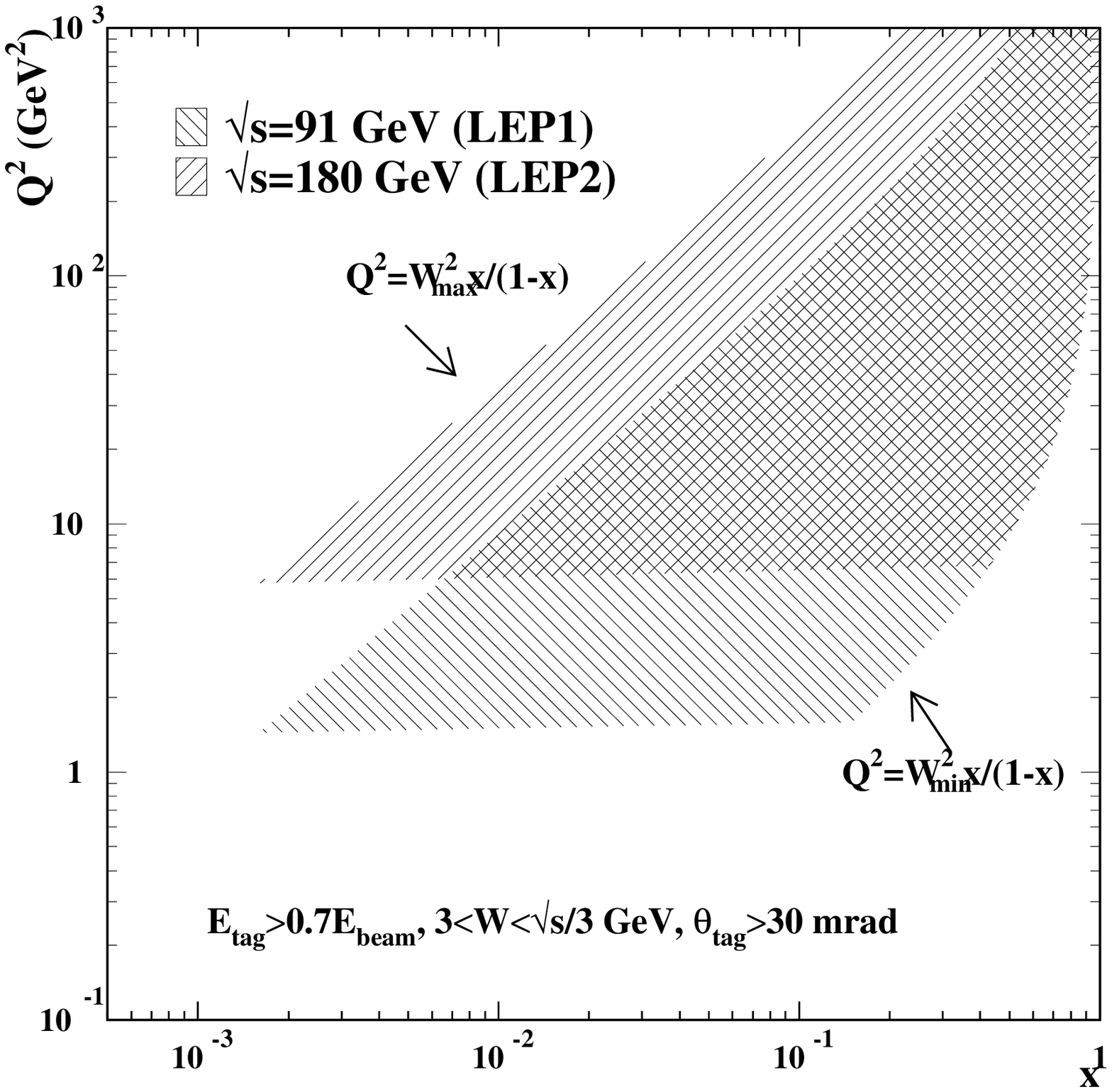,width=0.46\textwidth,height=6.5cm}
 &
\epsfig{file=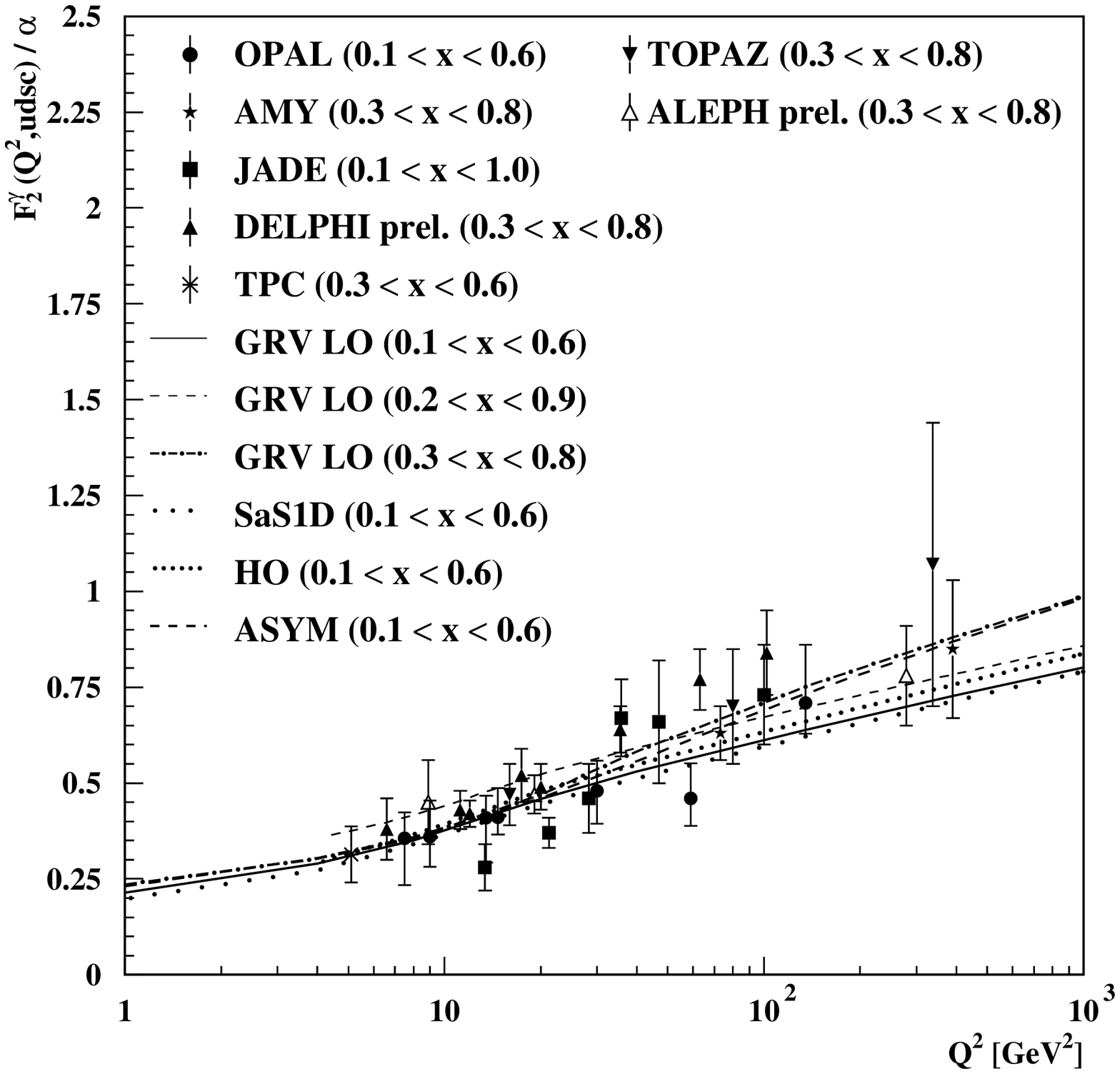,width=0.46\textwidth,height=6.5cm} 
\end{tabular}}
\put(6.1,1.1){(a)}
\put(13.5,1.1){(b)}
\end{picture}
\caption{\label{fig-cov}
a) Kinematical coverage of the ($Q^2,x$) plane at LEP1 and LEP2. 
b) The photon structure function $F_2^{\gamma}/\alpha$
as a function of $Q^2$.} 
\end{figure}

The photon structure function $F_2^{\gamma}(x,Q^2)$ can be measured at LEP
in the range $x>10^{-3}$ and $1<Q^2<10^3$~GeV$^2$ which makes
it possible to study the QCD evolution of $F_2^{\gamma}$ in a wide range
of $x$ and $Q^2$ (Fig.~\ref{fig-cov}a). 

The evolution of $F_2^{\gamma}$ with $\ln Q^2$ is
shown in Fig.~\ref{fig-cov}b using the currently available
$F_2^{\gamma}$ measurements for 4 active flavours. The data are compared to
the LO GRV~\cite{bib-grv} and the SaS-1D~\cite{bib-sas}
parametrisations, and to a higher
order (HO) prediction based on the NLO GRV parametrisation for
light quarks and on the NLO charm contribution calculated in
Ref.~12. The data are measured in different $x$ ranges. 
The comparison of the LO GRV curves
for these $x$ ranges shows that for $Q^2>100$~GeV$^2$ significant
differences are expected.
An augmented asymptotic prediction for $F_2^{\gamma}$ is also
shown. The contribution to $F_2^{\gamma}$ from the three light flavours is 
approximated by Witten's leading order asymptotic form~\cite{bib-witten}.
This has been augmented by adding a charm contribution 
evaluated from the Bethe-Heitler formula~\cite{WIT-7601}, and an estimate of 
the hadronic part of $F_2^{\gamma}$, which essentially 
corresponds to the hadronic part of the LO GRV parametrisation.
In the region of medium $x$ values studied here this asymptotic prediction 
in general lies higher than the GRV and SaS predictions but it is still 
in agreement with the data.
The importance of the hadronic contribution to $F_2^{\gamma}$
decreases with increasing $x$ and $Q^2$, 
and it accounts for only 15~\% of $F_2^{\gamma}$ at 
$Q^2= 59$~GeV$^2$ and $x = 0.5$.

As predicted by QCD the evolution of $F_2^{\gamma}$ leads
to a logarithmic rise with $Q^2$, but theoretical and experimental
uncertainties are currently too large for a precision test of perturbative QCD.
\begin{figure}[htbp]
   \begin{center}
      \mbox{
 \epsfig{file=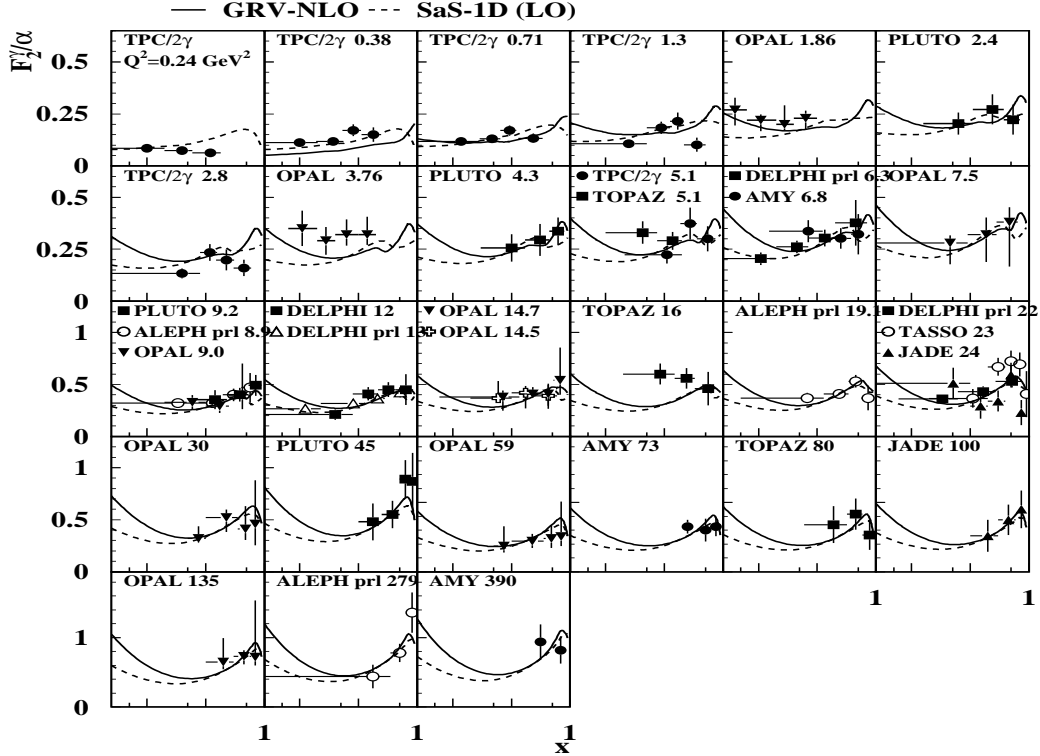,width=0.9\textwidth,height=10cm}
           }
   \end{center}
\caption{\label{fig-sf}
Measurements of the photon structure function $F_2^{\gamma}$
in bins of $x$ and $Q^2$. }
\end{figure}

\subsection{The photon structure function $F_2^{\gamma}$ at low $x$}
All currently available measurements~\cite{bib-f2g} of the photon
structure function are shown in Fig.~\ref{fig-sf}.
The data are compared to the NLO GRV~\cite{bib-grv} 
parametrisation and the LO SaS-1D~\cite{bib-sas} parametrisation. 
If the photon is purely hadron-like at low $x$,
a rise of the photon structure function
is expected in the low $x$ region for not too small 
$Q^2$, similar to the rise of the proton 
structure function observed at HERA. 
Only with the high statistics and high energy LEP2 data will it
be possible to access regions in $x$ and $Q^2$ where the
rise of $F_2^{\gamma}$ could be observed.
An interesting new low $x$ 
measurement of $F_2^{\gamma}$ is presented by OPAL
in the $x$ and $Q^2$ ranges $2.5\times10^{-3}<x<0.2$ and $1.1<Q^2<6.6$~GeV$^2$.
The measurement is consistent with a possible rise within
large systematic errors. It should  be noted that
the OPAL points are significantly higher than the previous
measurement by TPC/2$\gamma$ in a similar kinematic range.

\begin{wrapfigure}[20]{r}{2.7in}
\epsfig{file=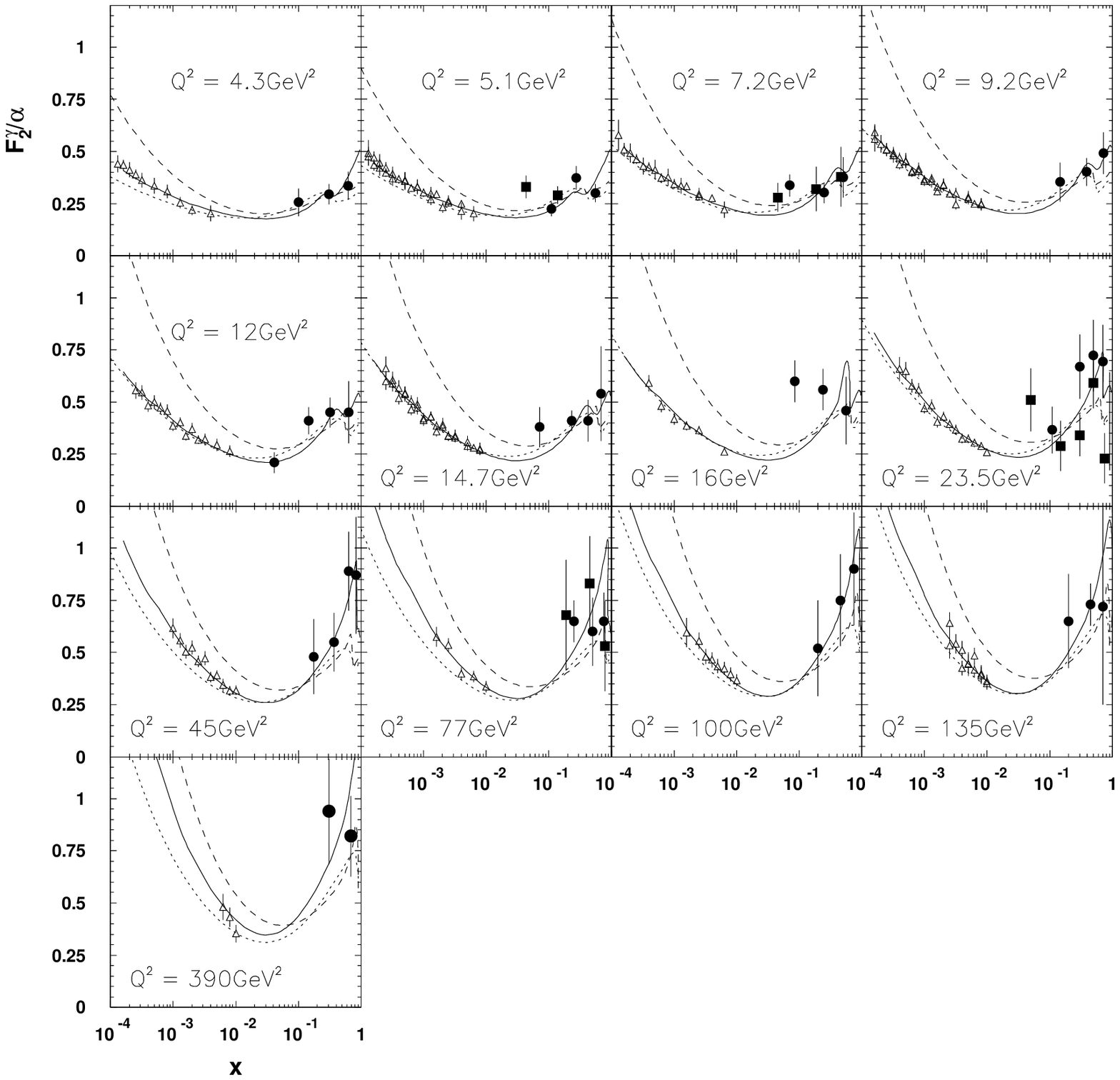,width=2.7in} 
\caption{\label{fig-levy}
The GAL parametrisation (continuous line)  compared to the $F_2^{\gamma}$ data
and the scaled $F_2^{\rm p}$ data using Gribov factorisation.
The dashed line is the LO GRV parametrisation
and the dotted line the SaS parametrisation.}
\end{wrapfigure}
In the kinematic region of the LEP measurements, the effect
of the $P^2$ evolution can become quite important. The 
ALEPH and OPAL $F_2^{\gamma}(x,Q^2)$ measurements have all not been corrected
for this effect, i.e. they are actually measurements
of $F_2^{\gamma}(x,Q^2,P^2)$. OPAL estimates that the effect
of the non-zero virtuality $P^2$ could increase
their low $x$ 
measurement of $F_2^{\gamma}$ by about 10~\% for  $P^2=0$.
This estimate is based on
the $P^2$ dependent parametrisation of the parton distributions by Schuler and 
Sj\"ostrand~\cite{bib-virsas}.

Since there is little experimental information about the low $x$
region, Gurvich, Abramowicz and Levy (GAL) have tried to use
the $F_2^{\rm p}$ data at low $x$ in order to constrain their 
new $F_2^{\gamma}$ parametrisation~\cite{bib-gal}. 
Gribov factorisation is assumed, based on the idea that
at high enough centre-of-mass energies all total hadronic cross-sections 
are dominated by an universal Pomeron trajectory which
allows to relate hadron and photon induced cross-sections.
Furthermore this factorisation assumption should also hold for virtual photons
at low $x$. 
The proton structure function $F_2^{\rm p}$ and the photon
structure function $F_2^{\gamma}$ are then related by
the total $\gamma$p and pp cross-sections:
$$ F_2^{\gamma}(x,Q^2)=
F_2^{\rm p}(x,Q^2)\frac{\sigma_{\gamma p}(W^2)}{\sigma_{pp}(W^2)}.$$
The total cross-section are obtained from a Donnachie-Landshoff 
parametrisation~\cite{bib-DL} of the data (see Sect.~\ref{sec-total}). 
This procedure is used to obtain pseudo-data for $F_2^{\gamma}$ at low $x$.
The evolution of the parton densities is done using a LO Altarelli-Parisi
evolution with a starting scale $Q_0^2=4$~GeV$^2$. The predicted
rise is much lower at low $x$ than predicted by the LO GRV
parametrisation which is also shown in Fig.~\ref{fig-levy}.

%\begin{wrapfigure}{r}{2.5in}
%\epsfig{file=/u/ws/soldner/ring/pfig6.eps,width=2.35in} 
%\caption{\label{fig-l3fafb} The structure functions $F^{\gamma}_B$
%and $F^{\gamma}_A$ for the process $\ee\rightarrow\ee\mu^+\mu^-$ (L3).
%The lines show the QED expectation.}
%\end{wrapfigure}
\subsection{Azimuthal correlations}
Only the structure function $F_2^{\gamma}$ has so far been determined
directly from measurements of 
double-differential cross-sections
for e$\gamma$ events with hadronic or leptonic final states.
It has been pointed out~\cite{bib-az} that azimuthal correlations 
in the final-state particles from two-photon collisions are
sensitive to additional structure functions.  
Azimuthal correlations can thus supplement the direct measurement of structure 
functions. ALEPH~\cite{bib-brew}, L3~\cite{bib-l3az} and OPAL~\cite{bib-azopal}
have measured azimuthal correlations using 
single-tag $\ee\rightarrow\ee\mu^+\mu^-$ events.

For single-tag events two independent angles can be defined
in the $\gamma\gamma^*$ centre-of-mass system 
assuming that the target photon direction is parallel
to the beam axis: The azimuthal angle $\chi$ is the angle between
the planes defined by the $\gamma\gamma^*$ axis and
the two-body final state and the $\gamma\gamma^*$ axis and the tagged electron.
The variable $\eta = \cos\theta^*$ is defined by the angle
$\theta^*$ between the $\mu^-$ and the $\gamma\gamma^*$ 
axis.

Neglecting the longitudinal component of the target photon,
%and setting $(1-y)$ to one, 
the cross-section can be written as
($F^{\gamma}_2 = 2xF^{\gamma}_T + F^{\gamma}_L$): 
\begin{equation}
\frac{{\rm d}\sigma ({\rm e\gamma \rightarrow e}\mu^+\mu^-)}
     {{\rm d}x {\rm d}y {\rm d}\eta {\rm d}\chi/2\pi} 
      \approx   \frac{2\pi\alpha^2}{Q^2} \left( \frac{1+(1-y)^2}{xy} \right) 
 \left[2x\tilde{F}^{\gamma}_T + 
                   \tilde{F}^{\gamma}_L -
                   \tilde{F}^{\gamma}_A \cos\chi +
                   \frac{\tilde{F}_B^{\gamma}}{2}\cos2\chi \right].  
\end{equation}
The conventional structure functions are recovered by integration over 
$\eta$ and $\chi$: $F_i^{\gamma}= \int_{-1}^{1} 
\int_0^{2\pi}\frac{d\chi d\eta}{2\pi} \tilde{F}_i^{\gamma} (i=2,A,B)$.
The structure functions $F_T$ and $F_L$ are related to the scattering of 
transverse and longitudinally polarized
virtual photons on a transverse target photon, respectively. 
$F_A^{\gamma}$ is related to the
interference terms between longitudinal and transverse
photons and $F_B^{\gamma}$ to the interference term between purely
transverse photons.
The longitudinal structure function \fl\ has been shown to be equal to the 
structure function \fb\ in leading order and for massless muons, 
although coming from different helicity states of the photons.  

\unitlength1cm
\begin{figure}[htbp]
\begin{picture}(15.0,5.8)
\put(0,2.9){\begin{tabular}{ccc}
\epsfig{file=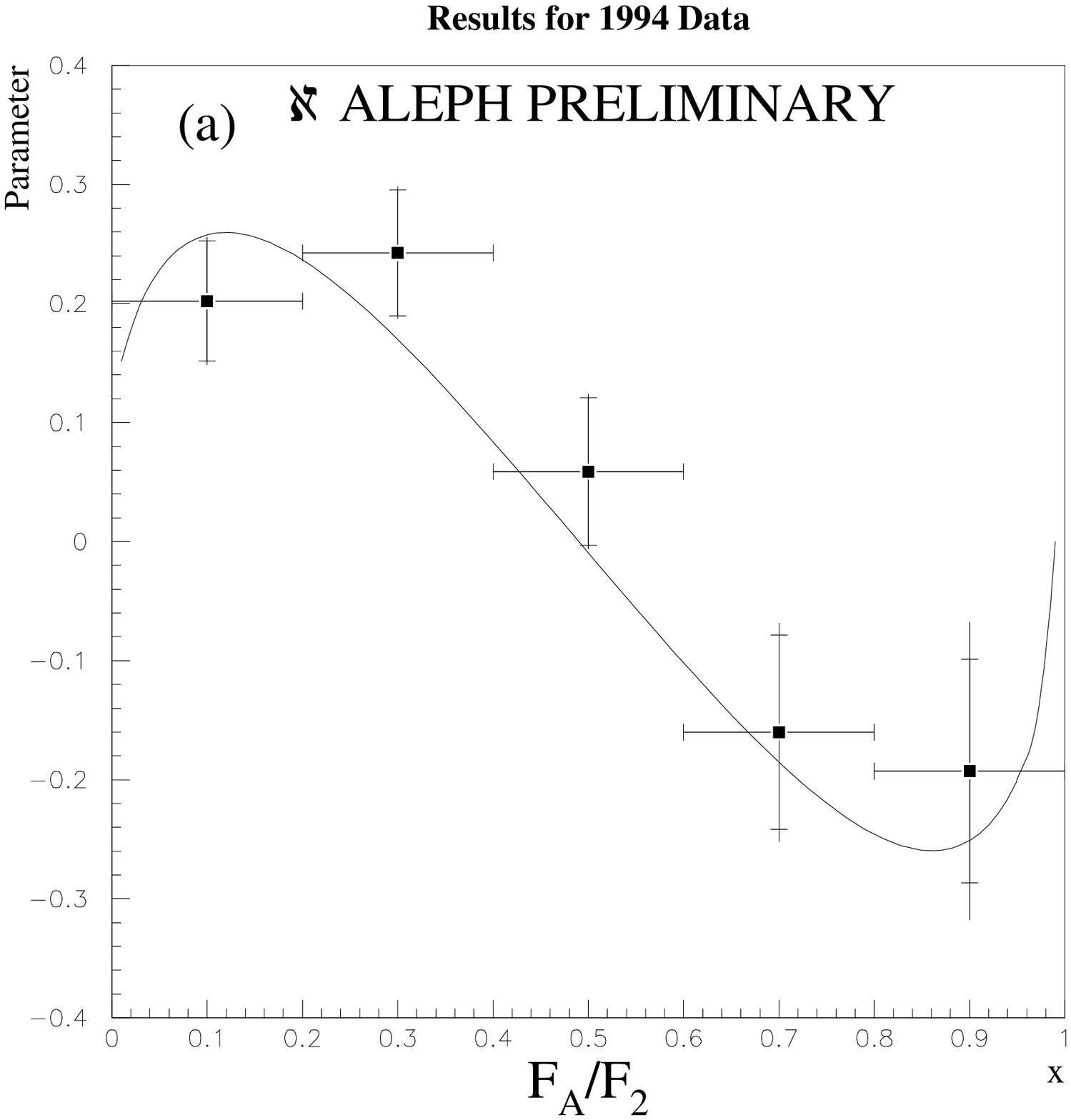,width=0.305\textwidth,height=5.8cm}
 &
\epsfig{file=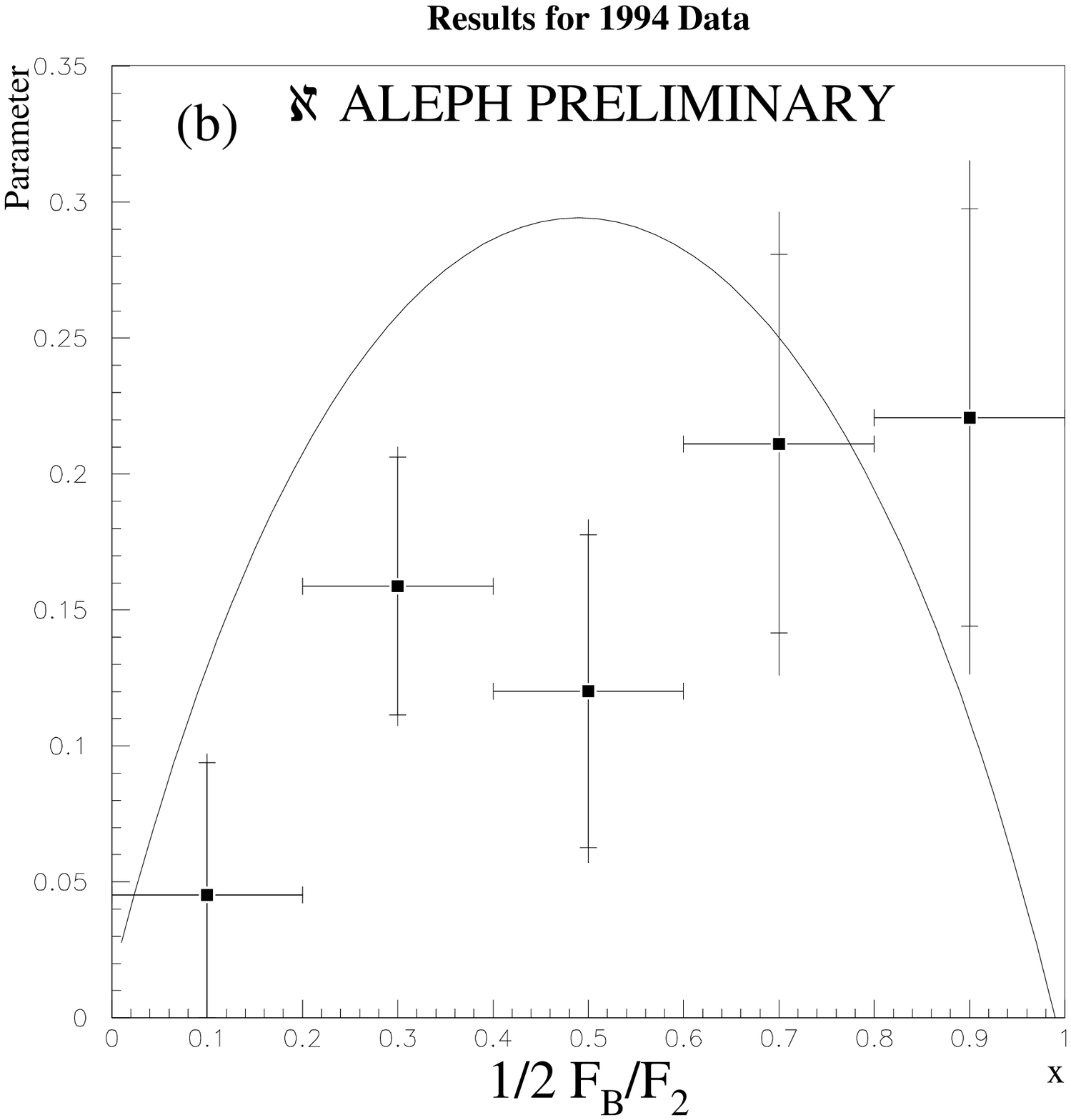,width=0.305\textwidth,height=5.8cm} 
 &
\epsfig{file=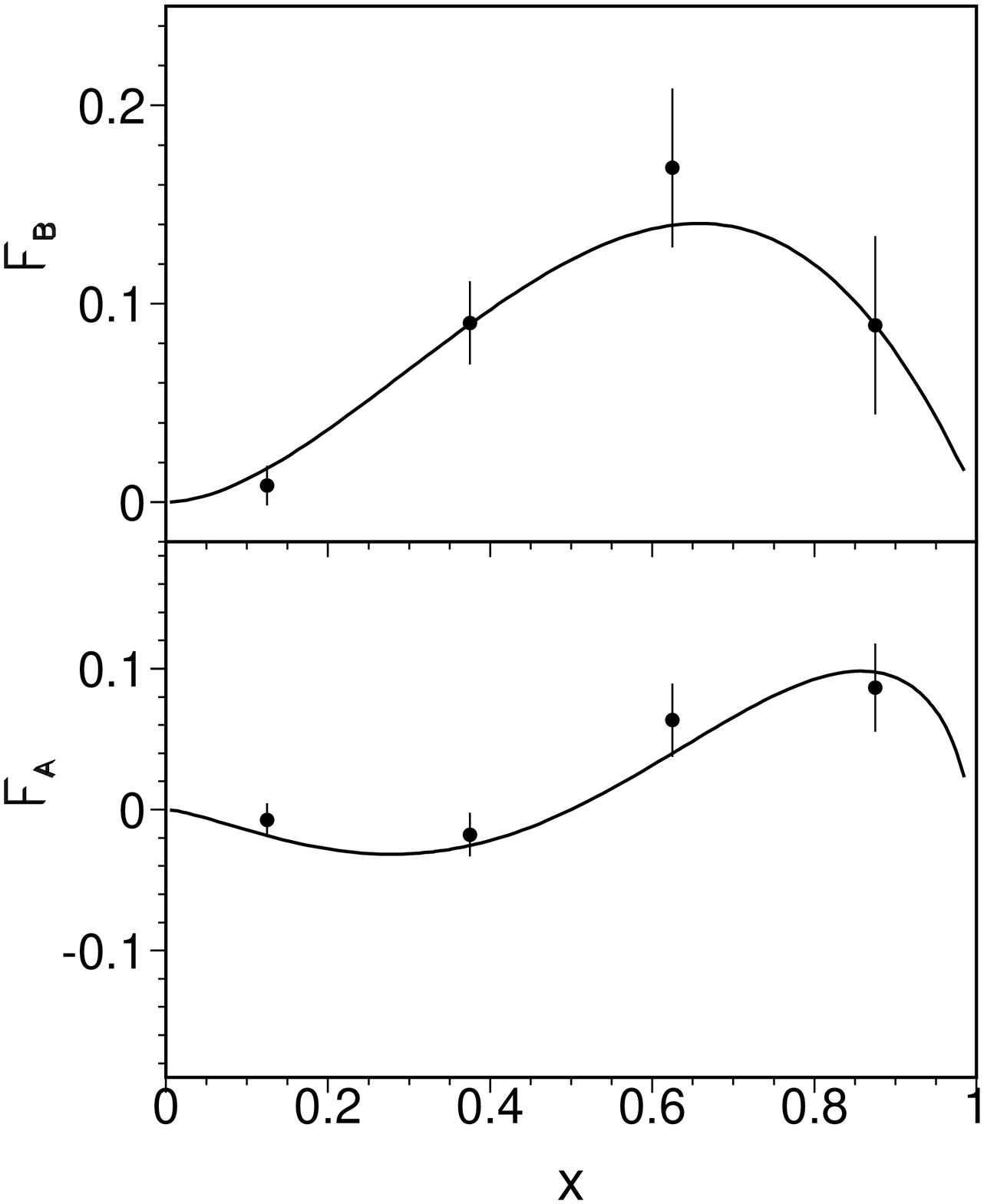,width=0.29\textwidth,height=5.8cm} 
\end{tabular}}
\put(12.0,4.9){L3}
\end{picture}
\caption{\label{fig-brew} The ratio of structure functions, 
$F^{\gamma}_A/F_2^{\gamma}$ and $1/2\cdot F^{\gamma}_B/F_2^{\gamma}$, 
for the process $\ee\rightarrow\ee\mu^+\mu^-$.
The lines show the QED expectation. The ALEPH measurement is not
corrected for the cut $|\eta|<0.8$.
The $Q^2$ ranges are $\langle Q^2\rangle=8.8$~GeV$^2$ (ALEPH) and
$1.4<Q^2<7.6$~GeV$^2$ (L3).}
\end{figure}
The variation of $F_A^{\gamma}$ and $F^{\gamma}_B$ with $x$ is 
in general consistent with QED (Fig.\ref{fig-brew} and 
Ref.~21).  
The measured values are significantly different from zero. 
Apart from being an interesting test of QED, these results
are especially important as a first step towards measuring
the additional structure functions for hadronic events
using azimuthal correlations. Such a measurement will be
much more difficult due to the problems related to
the jet finding in hadronic events.

\section{Jet production in \boldmath
$\gg$ and $\gamma$p scattering \unboldmath}
\label{sec-jet}
If the virtuality $Q^2$ of the probing photon in ep scattering or
the virtualities $Q^2$ and $P^2$ of both interacting photons 
in $\ee$ scattering are approximately zero, 
the photons can be considered to be quasi-real.
HERA can then be used as photon-proton collider with $\gamma$p 
centre-of-mass energies in the approximate range $40<W_{\gamma p}<300$ GeV and
LEP2 as a photon-photon collider with $\gg$ centre-of-mass energies
in the approximate range $10<W_{\gg}<120$~GeV. 
Anti-tagged $\gamma$p events at HERA and anti-tagged $\gg$ events
at LEP have a median $Q^2$ of about 10$^{-4}$ GeV$^2$ for
$Q^2<4$~GeV$^2$. The anti-tagging
condition is fulfilled if no scattered electron was found in the
main calorimeters. Events with $Q^2<10^{-2}$~GeV$^2$ 
are tagged using small angle calorimeters.

\begin{figure}[htb]
   \begin{center}
     \mbox{
          \epsfxsize=0.65\textwidth
          \epsffile{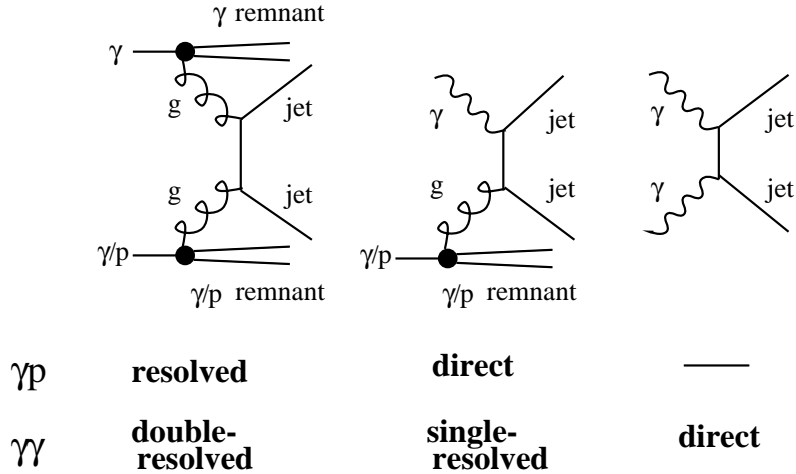}
           }
   \end{center}
\caption{Examples for leading order processes in $\gg$ and $\gamma$p
interactions.}
\label{fig-process}
\end{figure}

In LO different event classes can be defined
in $\gg$ and $\gamma$p interactions. The photons can either
interact as bare photons (``direct'') or as hadronic fluctuation
(``resolved''). The events are classified by the photon
interaction which leads to the nomenclature shown in Fig.~\ref{fig-process}.
In NLO the separation into event classes seizes to be unique,
since it depends on the factorisation scale for the photon, and
only the sum of the cross-section for the different event classes
is well defined.

\begin{figure}[htbp]
\begin{picture}(15.0,5.8)
\put(0,0.0){
%   \begin{center}
     \mbox{
          \epsfxsize=0.7\textwidth
          \epsffile{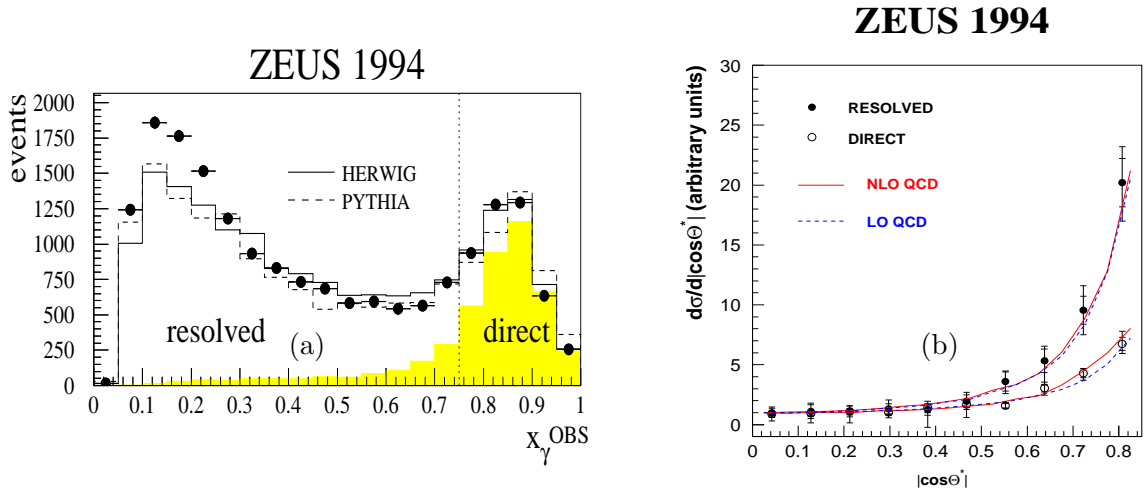}
           }
%   \end{center}
}
\put(4.0,1.5){(a)}
\put(12.4,1.5){(b)}
\end{picture}
\caption{\label{fig-theta} 
a) Uncorrected $x_{\gamma}^{\rm obs}$ distribution for 
$\gamma$p dijet events.
b) dijet angular distribution ${\rm d}\sigma/{\rm d}|\cos\theta^*|$
in the jet-jet centre-of-mass frame.}
\end{figure}

Within the LO picture we expect two hard
jets, i.e.~with large transverse energy $\ETJET$, in the final state
in addition to possible photon and/or proton remnants. 
These jets are related to the underlying parton
dynamics and can therefore be used to constrain
the structure of the colliding particles, photons and protons.
Direct and resolved events can be separated by using
the fraction $x_{\gamma}$ of the photon's momentum participating in the 
hard interaction. The direct events are expected to have $x_{\gamma}=1$. 
For $\gamma$p events a separation of the event classes has been achieved
experimentally by defining the variable~\cite{bib-xg}
$$x_{\gamma}^{\rm obs}=\frac{\sum_{\rm jets}\ETJET e^{-\etajet}}{2yE_{\rm e}},$$
with  $E_{\rm e}=27.5$~GeV being the electron beam energy.
The sum runs over the two jets with highest $\ETJET$ and
$E_{\gamma}=yE_{\rm e}$ is the energy of the initial state photon.
For two-jet final states, this observable is equivalent to the LO
definition of $x_{\gamma}$.
Fig.~\ref{fig-theta}a shows the uncorrected $x_{\gamma}^{\rm obs}$ distribution
measured by ZEUS in comparison to the LO MC models PYTHIA and
HERWIG. The peak at high $x_{\gamma}^{\rm obs}$ due to the
direct events is observed in the data. However, it is smeared out due
to hadronisation, QCD radiation and detector effects.

The measurement of jet cross-sections allows comparisons
with perturbative QCD calculations which are based on the matrix
elements $M(\cos\theta^*)$ for all possible parton scattering processes. 
The various matrix elements depend on the parton scattering angle $\theta^*$
in the parton-parton centre-of-mass frame.
A ZEUS measurement of the different angular distributions is shown
in Fig.~\ref{fig-theta}b. The $\cos|\theta^*|$
distribution is here given in the jet-jet centre-of-mass
system for resolved ($x_{\gamma}^{\rm obs}<0.75$) and direct dijet 
events ($x_{\gamma}^{\rm obs}>0.75$)~\cite{bib-theta}. 
The dijet invariant mass of these events is larger than 23 GeV.
The dominant LO diagrams for direct $\gamma$p processes involve
quark (fermion) exchange whereas those in resolved processes
involve gluon (boson) exchange. This leads to different
angular distributions which agree well both with the LO
QCD curves and with the NLO calculations by Harris and Owens~\cite{bib-harris}.
Similar results for $\gg$ scattering were reported by OPAL~\cite{bib-opalgg2}.

As we have seen, jet production is sensitive to the
parton content of the photon. 
In the kinematic range covered by $\ee$ experiments
the $F_2^{\gamma}$ measurements are mainly probing
the quark content of the photon, since the gluon distribution
only enters in higher order\footnote{Contrary to the proton case,
there exists no simple sum rule for the gluon distribution in the photon.
A sum rule has been derived by Frankfurt and Gurvich~\cite{bib-MSR1}
and Schuler~\cite{bib-MSR2}.}.
This is different in 
$\gg$ and $\gamma$p jet production where a large fraction of
the events are due to gluon initiated processes. 
Measurements of jet production cross-sections can therefore be used
to constrain the relatively unknown gluon distribution in the photon.

Different groups have followed different philosophies for extracting
information about the parton content of the photon from
jet cross-sections. In the first approach hadronic jet cross-sections
are measured and compared to calculations which use
different parametrisations of the photon's parton densities
as input. In the second approach, which is mainly followed
by H1, LO parton densities are extracted
from the measurements.

For the comparison of theory and experiment 
it is very important to define suitable
jet algorithms. Measurements of
jet cross-section in $\gg$ and $\gamma$p interactions
are usually made with cone jet finding 
algorithms~\cite{bib-cone}, since
resolved photon interactions are similar to hadron-hadron interactions.
A problem appears for overlapping jets which can be merged in iterative cone
algorithms. This is not described by a NLO calculation
which only contains three final state partons~\cite{bib-rsep}.

Other effects like hadronisation or the underlying event in 
resolved events are also not contained
in the theory. Underlying event is a loose term for
additional activity which, for example, can be caused by soft or 
hard interactions  of the photon and proton remnants of the same event.
The influence of these effects has to be studied
in detail before statements about the parton distributions
can be made, since the additional energy in the event directly affects
the jet cross-sections. The Monte Carlo models treat this effect
by introducing multiple parton interactions (MI).

\subsection{Jet shapes}
The internal structure of jets was studied by ZEUS~\cite{bib-shape}
by measuring jet shapes in photoproduction.
The jet shape is defined as the average fraction of $\ETJET$
that lies inside an inner cone of radius $r$ concentric with
the jet defining cone:
$$\psi(r)=
\frac{1}{N_{\rm jets}}\sum_{\rm jets}\frac{E_{\rm T}(r)}{E_{\rm T}(r=R)},$$
where $E_{\rm T}(r)$ is the transverse energy within the inner cone
of radius $r$ and $N_{\rm jets}$ the total number of jets.
\unitlength1cm
\begin{figure}[htbp]
\begin{picture}(15.0,6.5)
\put(0,3.25){\begin{tabular}{cc}
\epsfig{file=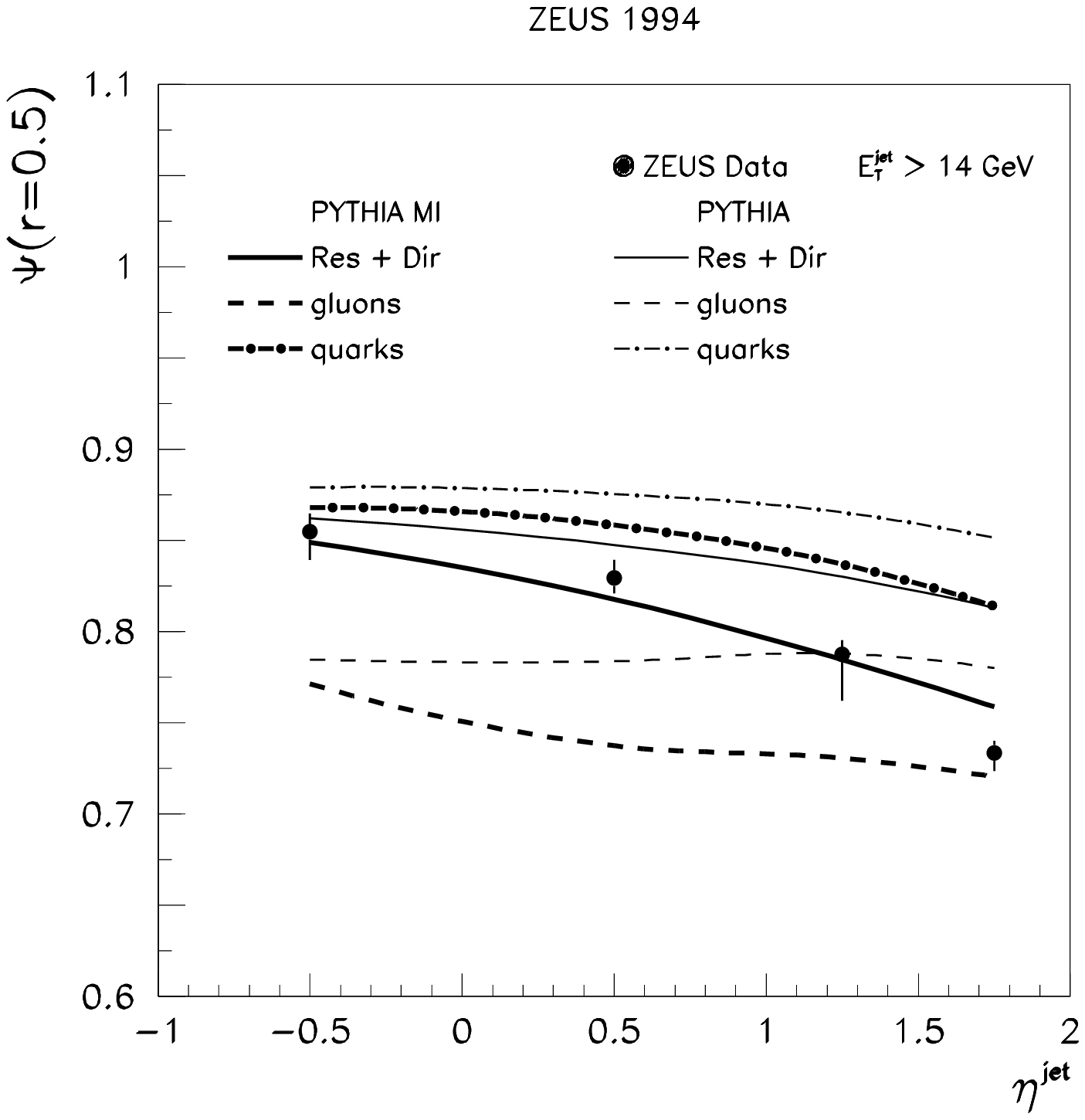,width=0.46\textwidth,height=6.5cm}
 &
\epsfig{file=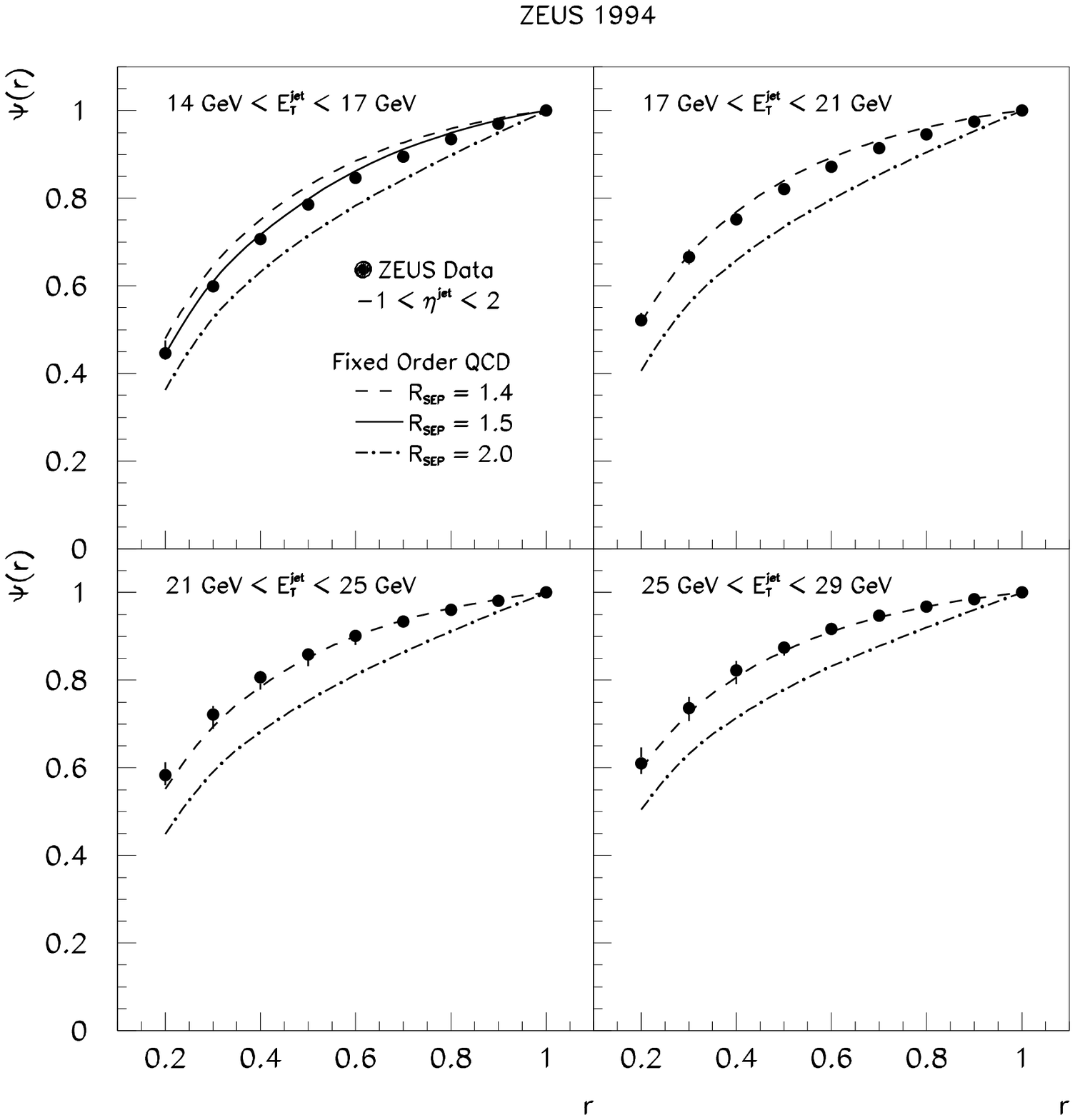,width=0.46\textwidth,height=6.5cm} 
\end{tabular}}
\put(1.6,1.4){(a)}
\put(9.0,1.4){(b)}
\end{picture}
\caption{\label{fig-shape} 
a) Jet shape $\psi(r)$ at a fixed value $r=0.5$ as a function of $\etajet$
for jets with $\ETJET>14$~GeV;
b) jet shapes $\psi(r)$ as a function of the radius $r$
for jets with $-1<\etajet<2$ in different $\ETJET$ regions.
}
\end{figure}
The $\etajet$ dependence of the jet shapes is compared
at a fixed value for $r$, $\psi(r=0.5)$, to the
PYTHIA prediction with and without multiple interactions
(MI) in Fig.~\ref{fig-shape}a. 
%In the MC generator
%parton radiation is necessary to reproduce
%the observed jet shapes.
The larger colour charge of gluon jets leads to increased
QCD radiation which broadens the jets, an effect
which is well known from jet production in $\ee$ scattering.
In the forward direction (positive $\etajet$)
the fraction of gluon jets in the MC increases, but also
the influence of MI increases. Both effects are
expected to lead to the broadening of the jets at high $\etajet$
which is observed in the data. In Fig.~\ref{fig-shape}b 
the jet shapes $\psi(r)$ for different $\ETJET$ ranges are 
also compared to fixed-order perturbative QCD
calculations~\cite{bib-rsep,bib-nloshapes}. 
Jets get narrower with increasing $\ETJET$. 
A new parameter $R_{\rm sep}$ was introduced
in the calculation to mimic the effects of overlapping and merging
of jets mentioned in the previous section. Two partons are not
merged into a single jet if their separation in the $\eta\phi$ plane
is more than $R_{\rm sep}$. This problem only exists
for the iterative cone algorithms used here, whereas
the $k_{\rm T}$ clustering algorithm~\cite{bib-ktclus} requires
no additional parameter $R_{\rm sep}$.
The choice $R_{\rm sep}=1.4$ gives a reasonable description of the
data, but at low $\ETJET$ a higher $R_{\rm sep}$ is needed.
This increase at low $\ETJET$ can be attributed to hadronisation effects and
a possible underlying event.

\subsection{Inclusive jet cross-sections and NLO calculations}
%The MC models used to describe $\gamma$p and $\gg$ collisions
%are all based on LO QCD. They also provide phenomenological models
%for effects like hadronisation, multiple interactions or
%the underlying event. These effects are not contained in the NLO
%calculations.
NLO jet cross-sections for $\gg$~\cite{bib-kleinwort,bib-aurenche} and 
$\gamma$p
interactions~\cite{bib-harris,bib-gpnlo,bib-kk} have been calculated by
many authors. To calculate jet cross-sections in perturbative QCD a hard scale
is required in the event which is usually the transverse momentum $p_{\rm T}$
of the final-state partons (or the jet). 
For the calculation it is assumed that the concept of factorisation can 
be applied. The LO jet cross-section is written as convolution of 
the parton density $f_{i/\gamma}$ of the photon and,
in the case of $\gamma$p scattering, the parton density 
$f_{j/p}$ of the proton with the LO matrix elements $M_{ij}$ for
 the
scattering of two partons $i$ and $j$:
%$$\frac{{\rm d}\sigma_{\rm ep}}{{\rm 
%d}x_{\gamma}{\rm d}x_{p(\gamma)}{\rm d}\cos\theta^*}
%\propto \sum_{ij} \frac{f_{i/\gamma}(x_{\gamma},p_{\rm T}^2)}{x_{\gamma}} 
%\frac{f_{j/p(\gamma)}(x_{p(\gamma)}, p_{\rm T}^2)}{x_{p(\gamma)}}\left | 
%M_{ij}(\cos\theta^*) \right |^2.$$
$$\frac{{\rm d}\sigma_{\rm ep(ee)}}{{\rm 
d}x_{1}{\rm d}x_{2}{\rm d}\cos\theta^*}
\propto \sum_{ij} \frac{f_{i/\gamma}(x_{1},p_{\rm T}^2)}{x_{1}} 
\frac{f_{j/p(\gamma)}(x_{2}, p_{\rm T}^2)}{x_{2}}\left | 
M_{ij}(\cos\theta^*) \right |^2.$$
The variables $x_1,x_2$ are either the momentum fractions
$x_{\gamma}, x_{\rm p}$ of the partons in the photon and the proton
(for $\gamma$p scattering) or the momentum fractions 
$x_{\gamma}^+,x_{\gamma}^-$ of the partons in the two interaction photons 
(for $\gg$ scattering).
In addition the photon flux from the electrons is 
taken into account using the Weisz\"acker-Williams Approximation.

The NLO correction term depends on the two factorisation scales,
the renormalisation scale and the cone size $R$.
NLO parton distributions of the photon should be used
for a consistent NLO calculation. 
Several such NLO parametrisations are available
GRV~\cite{bib-grv}, GS96~\cite{bib-GS96} and AFG~\cite{bib-AFG}.

%\begin{figure}[htbp]
%\begin{center}
%\begin{tabular}{cc}
%\begin{turn}{-90}
%\epsfig{file=/u/ws/soldner/lp97/write/sachs1.eps,width=5cm,
%height=0.46\textwidth}%\end{turn} &
%\begin{turn}{-90}
%\epsfig{file=/u/ws/soldner/lp97/write/sachs2.eps,width=5cm,
%height=0.46\textwidth} 
%\end{turn}
%\end{tabular}
%\end{center}
\begin{figure}[htbp]
\begin{tabular}{cc}
\epsfig{file=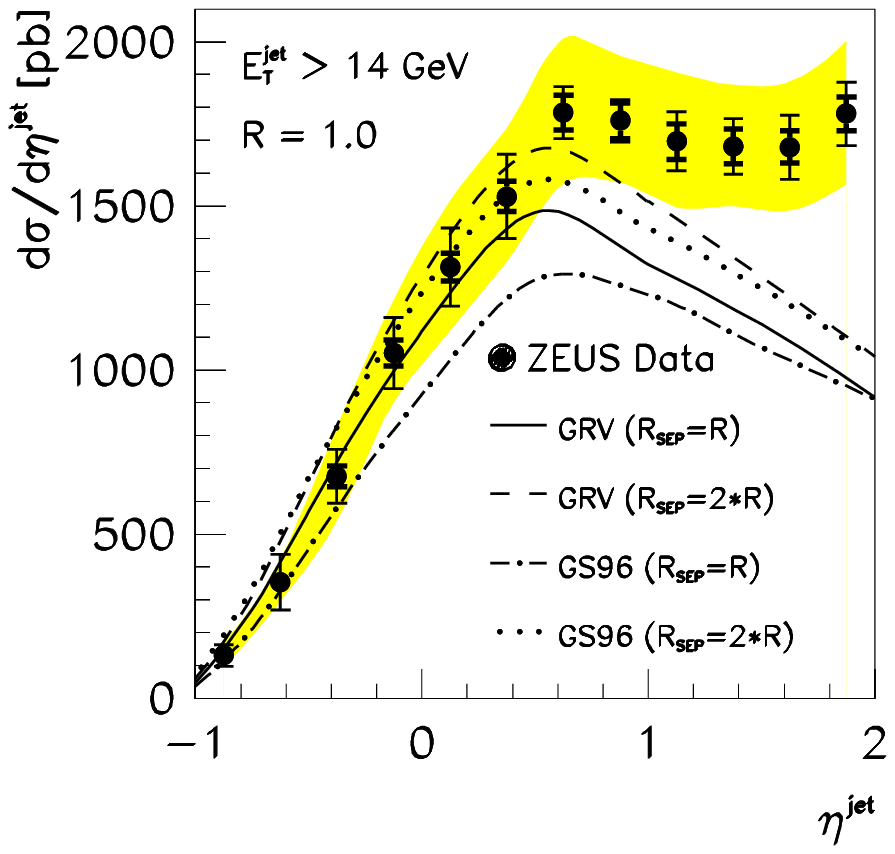,width=0.46\textwidth,height=6.0cm}
 &
\epsfig{file=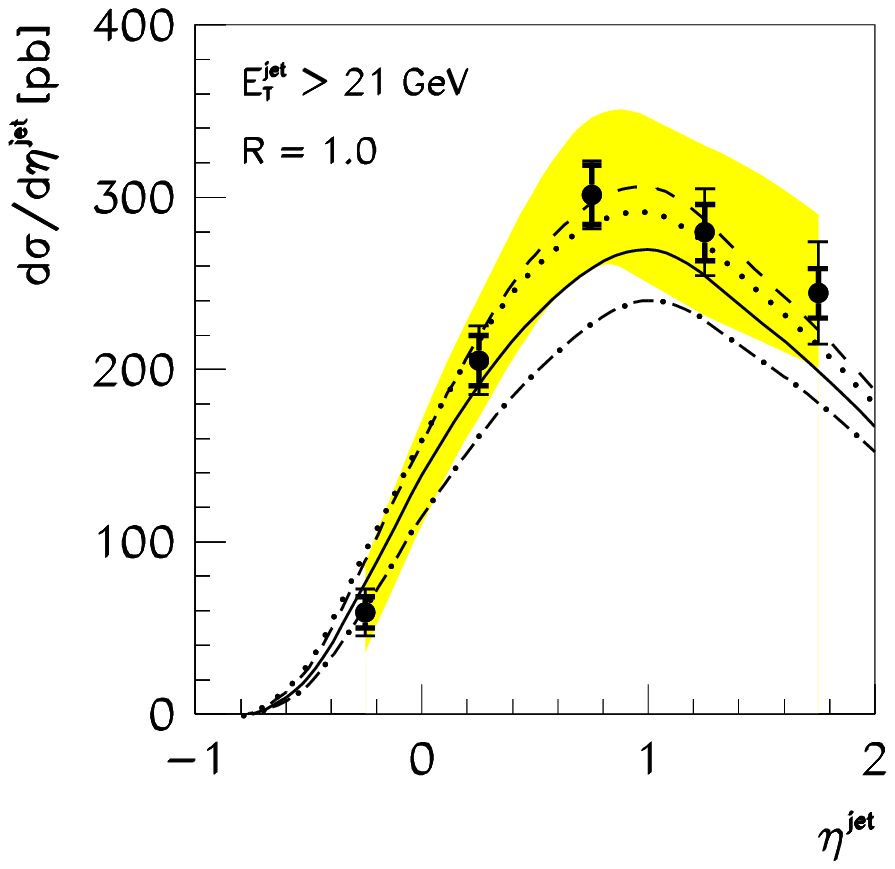,width=0.46\textwidth,height=6.0cm} 
\end{tabular}
\caption{\label{fig-zeusjet}The measured inclusive e$^+$p 
jet cross-section as a function
of $\etajet$ for jets with $\ETJET>14$~GeV and $\ETJET>21$~GeV
using a cone size $R=1$. The kinematic regime is defined
by $Q^2<4$~GeV$^2$ and $134<W<277$~GeV. The bands represent the
uncertainty due to the calorimeter energy scale. The curves
are the NLO calculations.
}
\end{figure}
ZEUS~\cite{bib-zeusjet} 
has compared measured inclusive single jet cross-sections in different
ranges of $\ETJET$ and $\etajet$ to the NLO calculation by 
Klasen and Kramer~\cite{bib-kk} (Fig.~\ref{fig-zeusjet}). The 
NLO GRV~\cite{bib-grv} and the the GS96~\cite{bib-GS96} parametrisations
are used for the photon. The discrepancies are large for
smaller $\ETJET$ in the forward  region
$\etajet>0.5$. This discrepancy disappears at higher $\ETJET$.
The agreement also improves if a smaller cone size is used
($R=0.7$) which corresponds to an effective increase
of the $\ETJET$ threshold compared to $R=1$. 
In the region of large discrepancy non-perturbative contributions from
the underlying event are expected to be large.

The NLO calculations are given for two values of $R_{\rm sep}$
($R_{\rm sep}=R, 2R$) which indicates part of the theoretical
uncertainty. 
The differences between NLO GRV and GS96 are of similar 
magnitude as the current theoretical and experimental
uncertainties. In addition ZEUS has also measured dijet~\cite{bib-zdijet}
and multi-jet~\cite{bib-zmjet} cross-sections.

\begin{figure}[htbp]
\begin{picture}(15.0,6.0)
\put(0,3.0){\begin{tabular}{cc}
\epsfig{file=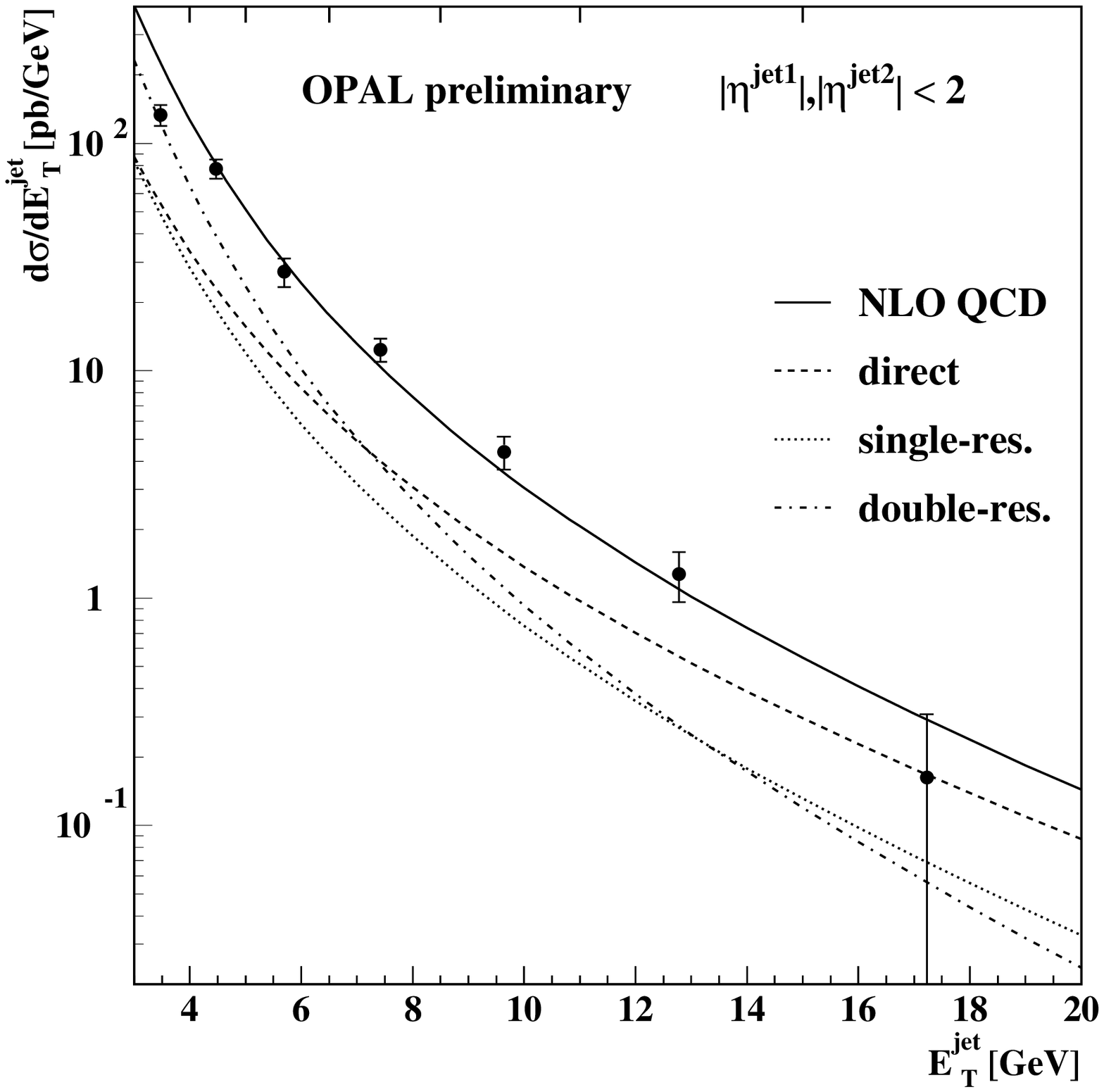,width=0.46\textwidth,height=6.0cm}
 &
\epsfig{file=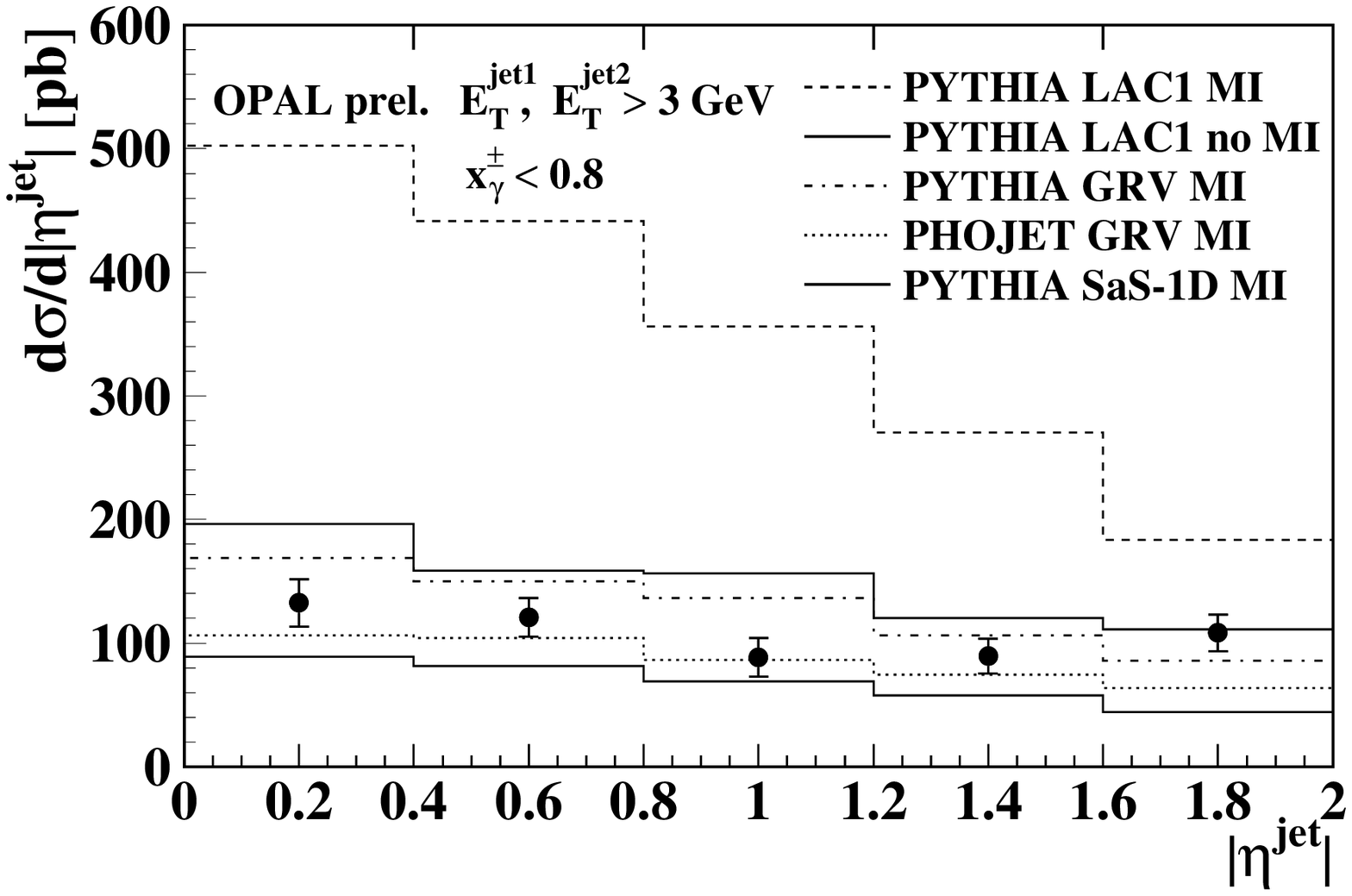,width=0.46\textwidth,height=6.0cm} 
\end{tabular}}
\put(1.5,3.4){(a)}
\put(9.0,3.4){(b)}
\end{picture}
\caption{\label{fig-ettwojet}
a) The inclusive $\ee$ dijet cross-section as a function
of $\ETJET$ for jets with $|\etajet|<2$ using a cone size $R=1$.
b) The inclusive dijet cross-section as a function
of $|\etajet|$ for jets in mainly double-resolved event
with $\ETJET>3$~GeV using a cone size $R=1$.}
\end{figure}

Inclusive one-jet and dijet cross-sections have also been
measured in $\gg$ scattering at an $\ee$ centre-of-mass energy of 
$\sqee=58$ GeV at TRISTAN~\cite{bib-amy,bib-topaz} and at $\sqee=130-172$ GeV
by OPAL~\cite{bib-opalgg,bib-opalgg2}. 
The $\ETJET$ distribution for dijet events in the range $|\etajet|<2$ 
measured by OPAL~\cite{bib-opalgg2} at $\sqee=161-172$~GeV
is shown in Fig.~\ref{fig-ettwojet}a. The measurements are
compared to a NLO calculation of the 
inclusive dijet cross-section~\cite{bib-kleinwort} 
which uses the NLO GRV parametrisation for the photon~\cite{bib-grv}.
The direct, single- and double-resolved parts and their sum are
shown separately. The data points are in good agreement with
the calculation except in the first bin where
theoretical and experimental uncertainties are large.

To study the sensitivity to the choice of parametrisation
for the parton distributions of the photon, OPAL has also measured
the inclusive dijet cross-section as a function of $|\etajet|$
for events with a large double-resolved 
contribution obtained by requiring $\xgpm<0.8$
(Fig.~\ref{fig-ettwojet}b).
The variables $\xgpm$ for the two incoming photons are 
defined in the same way as the $\gamma$p variable $x_{\gamma}$.
Ideally, for direct events without remnant jets $\xgp=1$ and $\xgm=1$,
whereas for double-resolved events both values $\xgp$ and $\xgm$
are expected to be much smaller than~1.

The inclusive dijet cross-section predicted by the two
LO QCD models PYTHIA~\cite{bib-pythia} and 
PHOJET~\cite{bib-phojet} differ significantly even if the same 
photon structure function (here LO GRV) is used. 
This model dependence reduces the
sensitivity to the parametrisation of the photon structure function.
Different parametrisations were used as input to the PYTHIA simulation. 
The LO GRV~\cite{bib-grv} and SaS-1D parametrisations~\cite{bib-sas} describe
the data equally well, but LAC1~\cite{bib-LAC1} which increases
the cross-section for gluon-initiated processes
overestimates the inclusive dijet cross-section
significantly. As in the case of $\gamma$p scattering a correct treatment
of multiple interactions is important. 
The PYTHIA cross-sections with and  without MI using LAC1
differ by more than a factor of two.

%For a more quantitative interpretation of these results
%in terms of parton distribution functions, it
%will be very important to understand the influence of
%multiple interactions on the jet cross-sections and
%to use jet definitions which will allow to compare
%theory (partons) and experiment (hadrons) directly. This
%is very similar to the problems of measuring jet cross-sections in 
%photoproduction at HERA discussed in these 
%proceedings~\cite{bib-butter,bib-klasen}. 

\subsection{Effective parton densities}
Following a procedure developed by Combridge and Maxwell~\cite{bib-comb},
H1 has measured the parton distributions of the photon, 
$f_{i/\gamma}$, using dijet events from $\gamma$p 
interactions~\cite{bib-h1eff}.
\begin{figure}[htbp]
\begin{tabular}{cc}
\epsfig{file=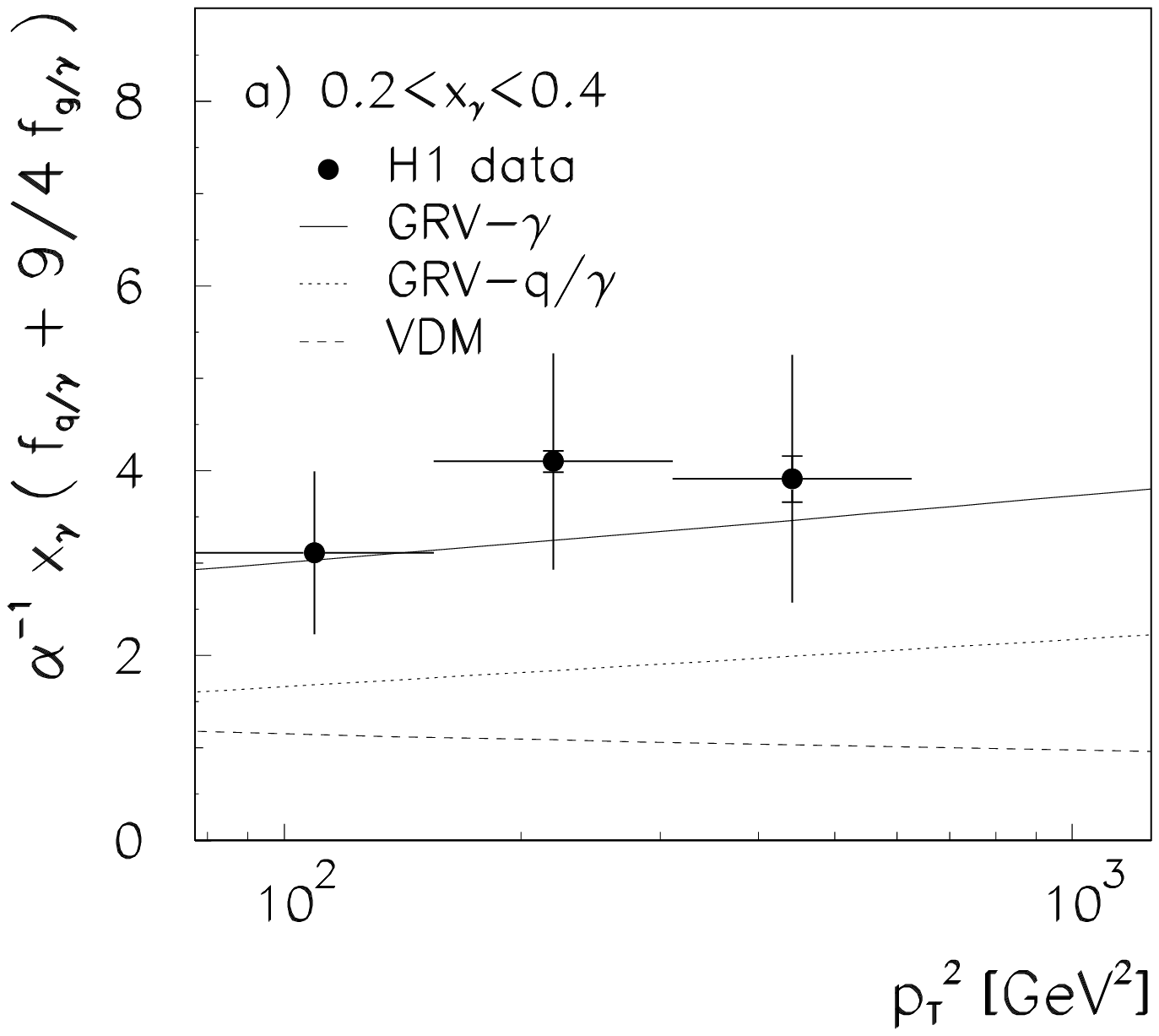,width=0.46\textwidth,height=6.5cm}
 &
\epsfig{file=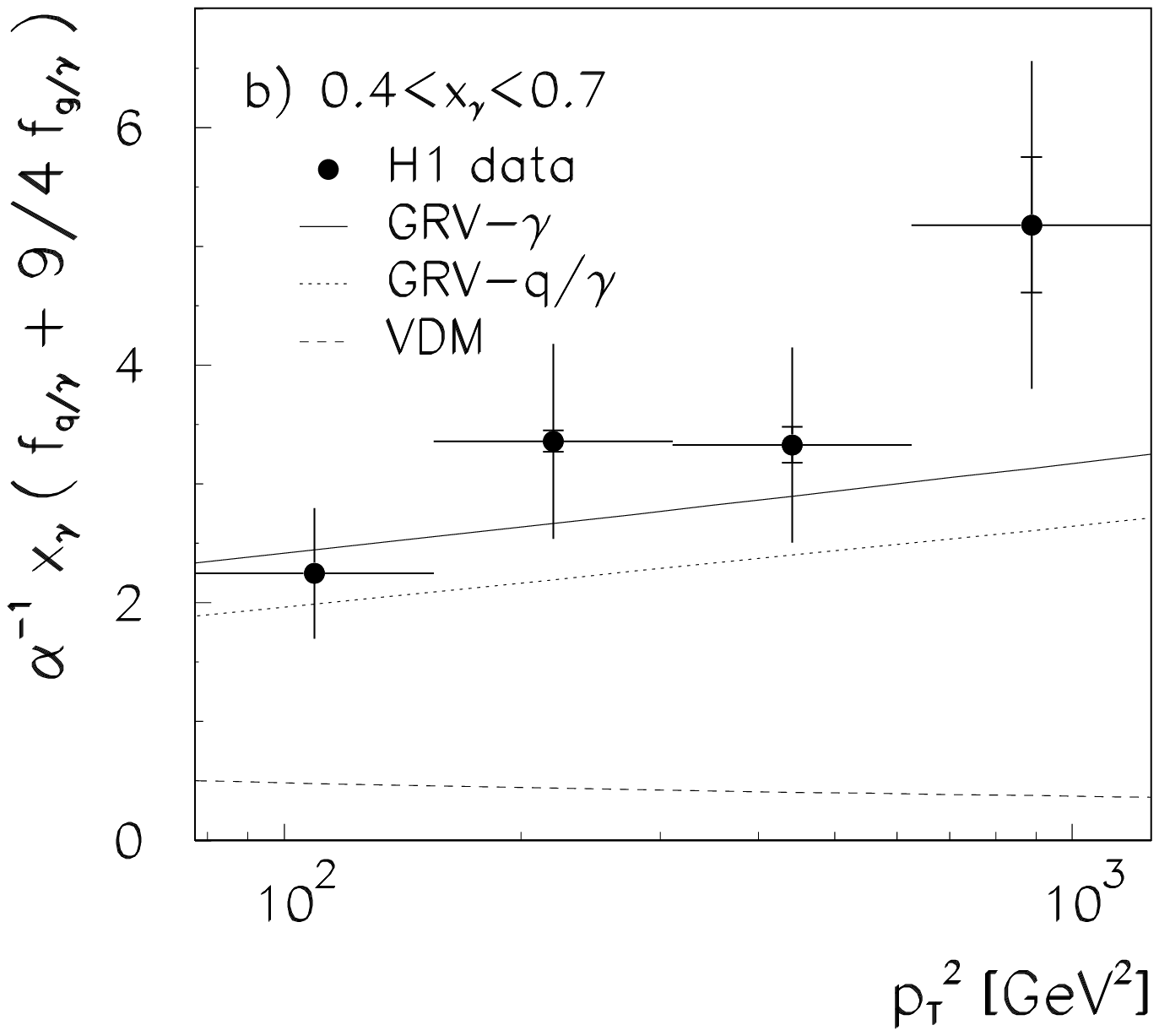,width=0.46\textwidth,height=6.5cm} 
\end{tabular}
\caption{\label{fig-eff} The LO effective parton distribution 
$\alpha^{-1}x_{\gamma}\tilde{f}_{\gamma}(x_{\gamma},p_{\rm T}^2)=
\alpha^{-1}x_{\gamma}(\tilde{f}_{q/\gamma}(x_{\gamma},p_{\rm T}^2)
+(9/4)\tilde{f}_{g/\gamma}(x_{\gamma},p_{\rm T}^2))$
as a function of the squared parton transverse momenta, $p_{\rm T}^2$.
}
\end{figure}
This method is based on LO matrix elements. It exploits the fact that
in the studied range $0.2<x_{\gamma}<0.7$ 
the dijet cross-section is dominated by the parton scattering
processes qq$'\rightarrow$qq$'$, qg$\rightarrow$qg and
gg$\rightarrow$gg. The shapes of the angular distributions of
the squared matrix elements $|M_{ij}(\cos(\theta^*)|^2$ is similar
for all these sub-processes and the rates only differ by the
ratio 9/4 of the colour factors. The matrix element can 
therefore be approximated by a ``Single Effective Subprocess'' 
matrix element, $M_{\rm SES}$,
and the parton densities are combined into effective parton density
functions:
$$\tilde{f}_{\gamma}(x_{\gamma},p_{\rm T}^2)
 \equiv \sum_{n_f}\left(f_{q/\gamma}(x_{\gamma},p_{\rm T}^2)
+f_{\overline{q}/\gamma}(x_{\gamma},p_{\rm T}^2)
\right)+\frac{9}{4}f_{g/\gamma}(x_{\gamma},p_{\rm T}^2)$$
$$\tilde{f}_{p}(x_{\gamma},p_{\rm T}^2)
 \equiv \sum_{n_f}\left(f_{q/p}(x_{\gamma},p_{\rm T}^2)+
f_{\overline{q}/p}(x_{\gamma},p_{\rm T}^2)
\right)+\frac{9}{4}f_{g/p}(x_{\gamma},p_{\rm T}^2)$$
where the sum runs over the number of quark flavours, $n_{\rm f}$,
and $p_{\rm T}$ is the transverse momentum of the final state parton.
The dijet cross-section is then replaced by 
$$\frac{{\rm d}\sigma_{\rm ep}}{{\rm 
d}x_{\gamma}{\rm d}x_{p}{\rm d}\cos\theta^*}
\propto \sum_{ij} \frac{\tilde{f}_{i/\gamma}(x_{\gamma},
p_{\rm T}^2)}{x_{\gamma}} 
\frac{\tilde{f}_{j/p}(x_{p}, 
p_{\rm T}^2)}{x_{p}}\left | 
M_{\rm SES}(\cos\theta^*) \right |^2.$$
The effective parton distribution $\tilde{f}_{\gamma}$ is extracted
from the data by measuring the double-differential dijet
cross-section d$^2\sigma/{\rm d}x_{\gamma}{\rm d}\log p_{\rm T}$.

The H1 measurements of the effective parton distributions 
in two different $x_{\gamma}$ ranges
($0.2<x_{\gamma}<0.4$ and $0.4<x_{\gamma}<0.7$) are shown Fig.~\ref{fig-eff}
as a function of $p_{\rm T}^2$. The $\log p_{\rm T}^2$ dependence
of $\tilde{f}_{\gamma}$ can now be compared to the QCD evolution
using the inhomogeneous DGLAP equations. This is done
using the LO GRV parametrisations~\cite{bib-grv} for the pion
and the photon. Assuming that the parton distribution of
the pion and the $\rho$ are similar, the purely hadronic (VDM) part
of the parton distribution functions is estimated by scaling
down the pion distribution with a VDM factor related
to the photon-$\rho$ conversion probability. The VDM picture
fails to describe the data. 

The LO GRV photon parametrisation
is constructed to be purely hadronic
at very low scales. Due to the pointlike $\gamma\rightarrow\qqbar$
term in the DGLAP evolution used to evolve the photon distributions
to large scales, one obtains the logarithmic rise observed
for the photon structure function $F_2^{\gamma}(x,Q^2)$ in e$\gamma$ scattering
(see Sect.~\ref{sec-egamma} and Fig.~\ref{fig-cov}b).
H1 compares the LO GRV distribution of the photon for all partons
and for quarks only to their data and both the rise and
the normalisation of the data are well reproduced. At lower
$x_{\gamma}$ the gluon distribution contributes about half, whereas
in the range $0.4<x_{\gamma}<0.7$ the quark distribution dominates
as expected.

\subsection{Leading order gluon densities}
The gluon density in the photon has been extracted
directly from data by H1 using a two-step procedure.
First the distribution of the momentum fraction $x_{\gamma}$ is 
unfolded from the data. In the second step the $x_{\gamma}$
distribution of the gluon initiated processes is obtained
by applying correction factors which are based on the
process definitions used in the Monte Carlo. Since
these process definitions are valid in LO, only LO
gluon densities can be extracted.

H1 has originally performed this measurement using dijet 
events~\cite{bib-h1gluon2}. This analysis suffers
from two large systematic uncertainties related to
the jet finding: the knowledge of the energy scale
of the H1 calorimeter and the effect of multiple interactions.
In a new analysis of the gluon density, H1 has
avoided these uncertainties by using events with high
transverse 
\begin{wrapfigure}{r}{0.45\textwidth}
\epsfig{file=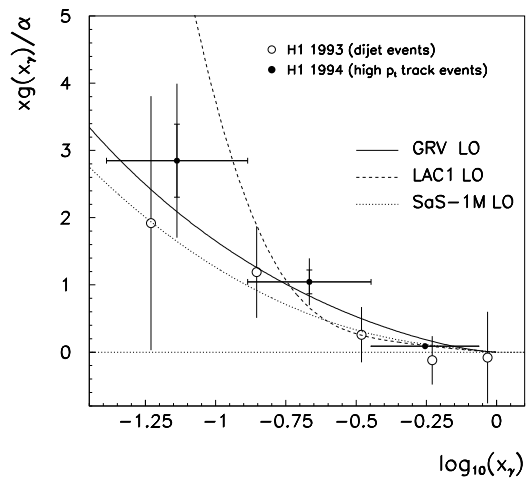,width=0.45\textwidth} 
\caption{\label{fig-h1gluon}
LO gluon distribution $x_{\gamma}g(x_{\gamma})/\alpha$ of the photon.
}
\end{wrapfigure}
momentum hadrons instead of dijet events~\cite{bib-h1gluon1}.
The price to be paid is a stronger sensitivity to
hadronisation effects. Events are required to
contain at least one charged particle with 
transverse momentum $p_{\rm T}>2.6$~GeV/$c$
and pseudorapidity $|\eta|<1$.
The variable $x_{\gamma}^{\rm rec}$ is reconstructed by summing over
all hadrons with $p_{\rm T}>2.0$~GeV/$c$ and $|\eta|<1$:
$$ x_{\gamma}^{\rm rec} = \frac{1}{E_{\gamma}}
\sum_{n} p_{\rm T}  e^{-\eta}$$
with $E_{\gamma}$ being the photon energy reconstructed from
the low $Q^2$ tagged electron ($Q^2<10^{-2}$~GeV$^2$).

After unfolding the LO gluon
distribution $x_{\gamma}g(x_{\gamma})/\alpha$ is 
shown in Fig.~\ref{fig-h1gluon}. 
Both analyses yield consistent results.
The gluon distribution rises at low $x_{\gamma}$, but
not as steeply as predicted by the LAC1 distribution~\cite{bib-LAC1}.
The average scale is given by the average squared transverse momentum
of the final state partons. It is 75~(GeV/$c$)$^2$ for the dijet analysis
and 38~(GeV/$c$)$^2$ for the hadron analysis.

\subsection{Virtual photon structure}
The measurement of the photon structure function
of virtual photons in $\ee$ collisions requires
the detection of both scattered electrons. 
Only PLUTO~\cite{bib-virpluto} has published such a measurement
for a virtuality $Q^2$ of the probing photon
of $5$~GeV$^2$ and a virtuality $P^2$ of the target
photon of $0.35$~GeV$^2$.
More double-tag measurements are to be expected
soon from the LEP2 data.

\begin{figure}[htbp]
\begin{picture}(15.0,6.5)
\put(0,3.25){\begin{tabular}{cc}
\epsfig{file=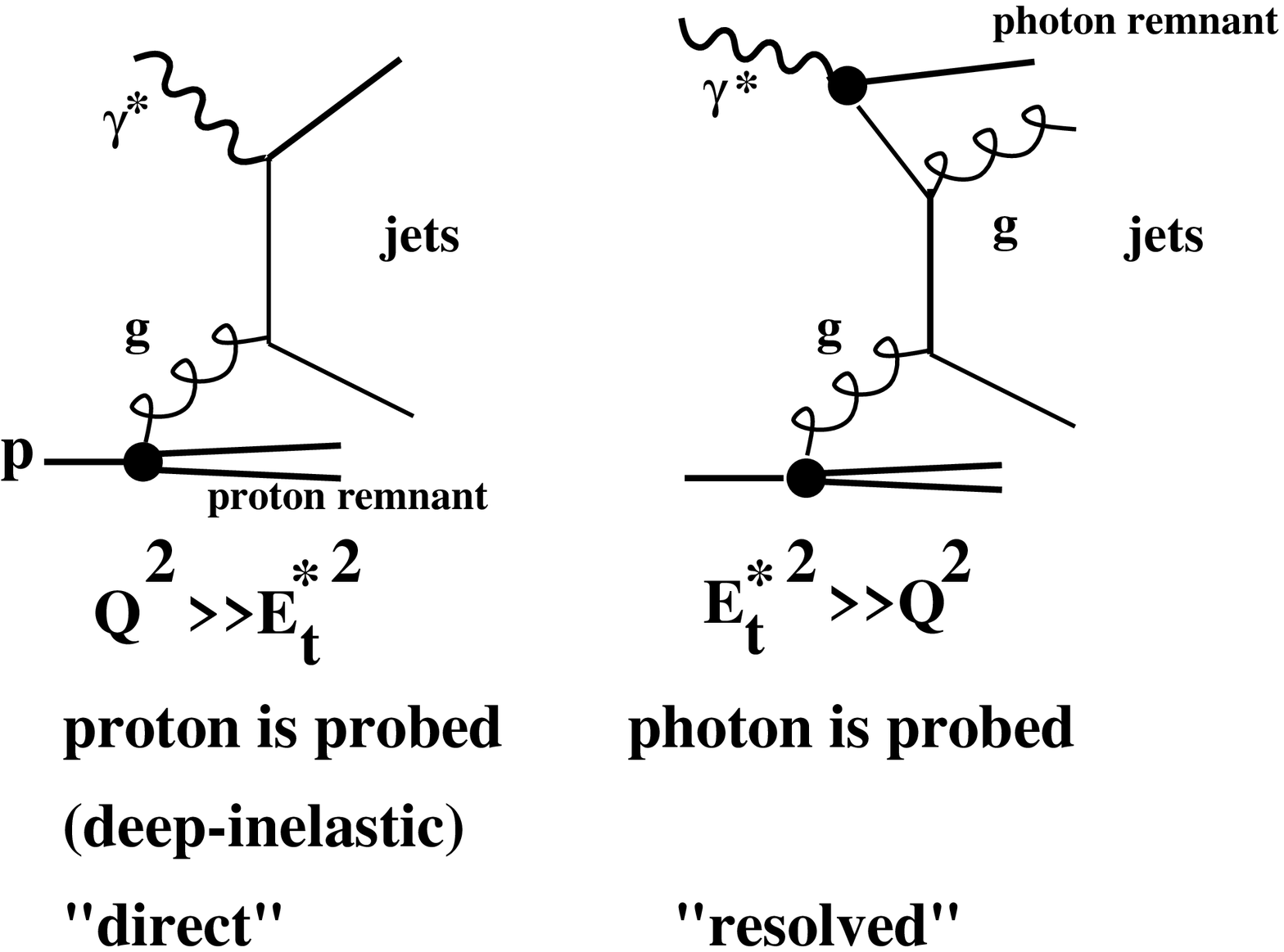,width=0.46\textwidth,height=6.5cm}
 &
\epsfig{file=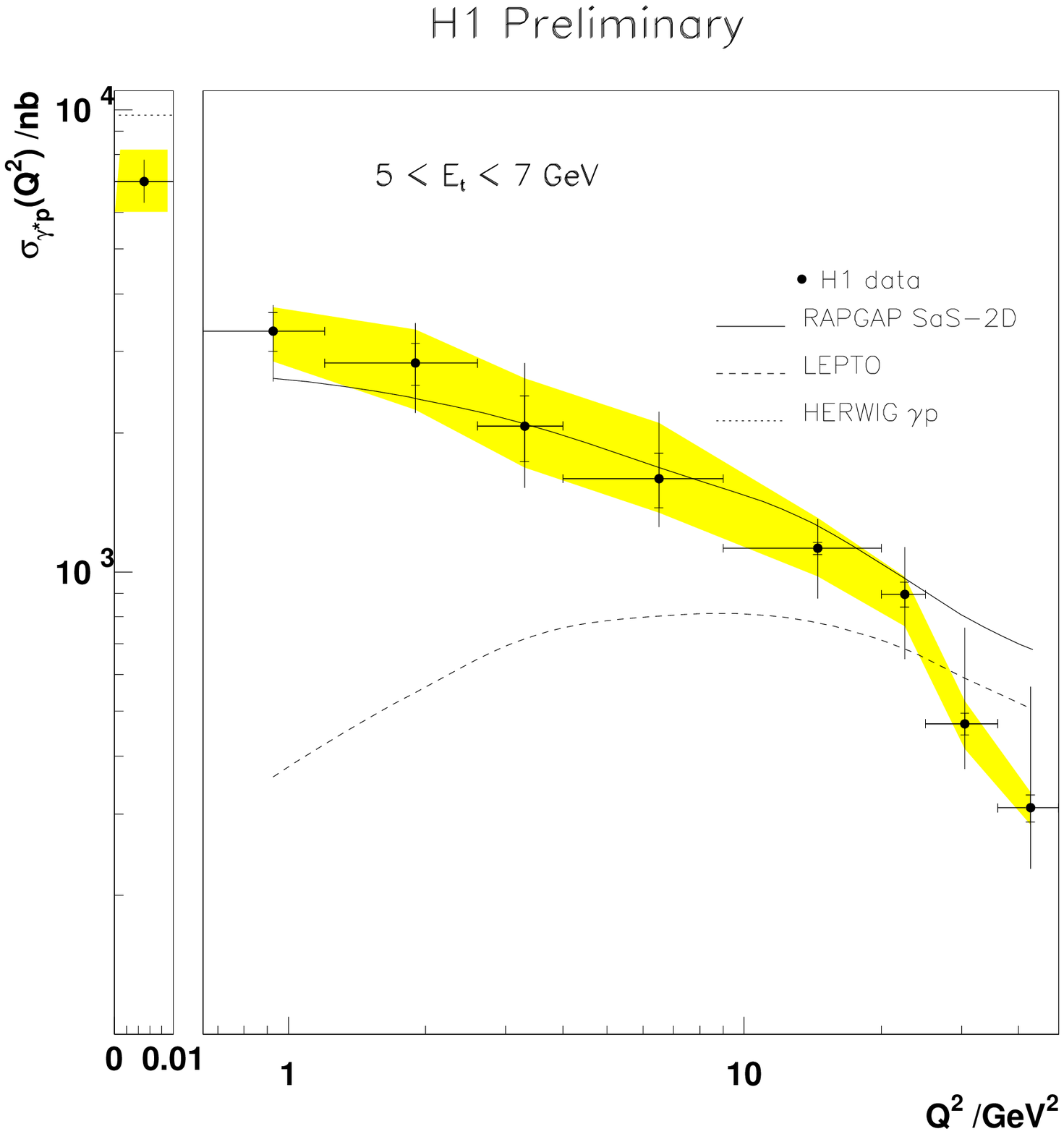,width=0.46\textwidth,height=6.5cm} 
\end{tabular}}
\put(0.3,5.5){(a)}
\put(9.4,5.5){(b)}
\end{picture}
\caption{\label{fig-h1vir}
a) Schematic drawing of the two kinematic regimes studied in this analysis.
b) Inclusive $\gamma^*$p jet cross-section as a function of $Q^2$
for jets with 
$5<\ESJET<7$~GeV and $-2.5<\eta^*<-0.5$ (in the $\gamma^*$p frame).}
\end{figure}

H1 has studied the structure of the virtual photon
by measuring the $Q^2$ dependence of jet production~\cite{bib-h1vir}.
This introduces a new scale in addition to
the transverse energy of the jet which
is the relevant scale for real photoproduction ($Q^2=0$).
In the kinematic range $Q^2>>(\ESJET)^2$ the ``classical''
picture of deep-inelastic scattering (DIS) can be applied
where the pointlike virtual photon probes the parton content
of the proton (Fig.~\ref{fig-h1vir}a).
The jet transverse energy $\ESJET$ is here calculated in the
$\gamma^*$p centre-of-mass system. 
However, a small fraction of the events
contains jets with $(\ESJET)^2>Q^2$. In this case
the process can be viewed as probing the structure of the virtual 
photon using the partonic structure of the proton.

H1 has measured the inclusive jet cross-section 
$\sigma_{\gamma^*\rm p}$ for the process $\gamma^*\mbox{p}
\rightarrow\mbox{jet}+X$ as a function of $Q^2$ 
for different $\ESJET$ ranges. In Fig.~\ref{fig-h1vir}b 
the cross-section $\sigma_{\gamma^*\rm p}(Q^2)$ is shown for jets with
$5<\ESJET<7$~GeV in comparison to various MC models.
LEPTO just simulates deep-inelastic scattering, i.e. it
contains no resolved photon processes. RAPGAP contains
both deep-inelastic scattering and resolved processes. Within RAPGAP
the SaS-2D parton distributions for the photon is used which include
a model for the $Q^2$ suppression for both the 
non-perturbative VDM and the perturbative anomalous part of 
the virtual photon.

The DIS picture completely fails to describe the data for
scales $Q^2<<(\ESJET)^2$, 
but it approaches the data for $Q^2\approx (\ESJET)^2\approx 36$~GeV$^2$,  
The RAPGAP model with SaS-2D parton
distributions is in good agreement with the data over the whole
$Q^2$ range. This observation is complemented by a measurement
of the fraction of the photon's energy which is assigned to
the photon remnant. This fraction decreases with increasing
$Q^2$ as expected if the resolved photon component is
suppressed with increasing $Q^2$.

\section{Prompt photon production}
Prompt photon production in $\gamma$p interactions at HERA is
dominated by the LO direct Compton process $\gamma\q\rightarrow\gamma\q$
and 
the resolved processes $\q\g\rightarrow\q\gamma$ and $\qqbar
\rightarrow\g\gamma$. ``Prompt'' means that these photons
are not produced in the fragmentation process or by particle
decays. The resolved process could be used to constrain
the quark content of the photon at 
\begin{wrapfigure}[15]{r}{0.45\textwidth}
\epsfig{file=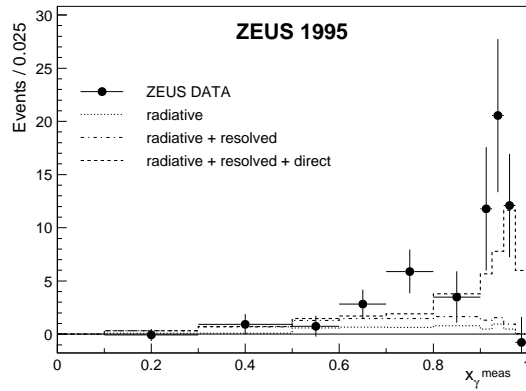,width=0.45\textwidth} 
\caption{\label{fig-prompt}
$x^{\rm meas}_{\gamma}$ distribution of prompt photons measured
by ZEUS.}
\end{wrapfigure}
medium $x_{\gamma}$, complementing 
the $F_2^{\gamma}$ measurements from e$\gamma$ scattering.
The advantage of having a clean final state without
hadronisation uncertainties is however largely compensated by
the small cross-section.

ZEUS~\cite{bib-prompt} has measured the production cross-section for
isolated photons with a transverse energy of $5\le E_{\rm T}^{\gamma}<10$~GeV
and in the pseudorapidity range $-0.7\le\eta^{\gamma}\le 0.8$,
in association with a jet of $\ETJET>5$~GeV in the range
$-1.5\le\etajet\le 1.8$.
Within a total error of more than 25~\%
the cross-section was found to be in good agreement with NLO calculations 
by Gordon~\cite{bib-gordon} using the GS and NLO GRV parton densities
for the photon. ZEUS has measured 
$x_{\gamma}$ by summing over the energies
$E$ and longitudinal momentum components $p_{\rm z}$ of the
photon and the calorimeter cells which are part of the jet
$$x^{\rm meas}_{\gamma}
=\frac{\sum_{\gamma,\rm jet}(E-p_{\rm z})}{2y_{\rm JB}E_{\rm e}}.$$
The variable $y_{\rm JB}=\sum_h(E-p_{\rm z})/2E_{\rm e}$ 
is calculated using the calorimeter cells associated to the 
final-state hadrons.
The $x^{\rm meas}_{\gamma}$ distribution of the ZEUS prompt photon
signal is shown in Fig.~\ref{fig-prompt}. A clear peak at 
$x^{\rm meas}_{\gamma}>0.8$ is observed. Using the LO Monte Carlo
generator PYTHIA about 75~\% of the events in this region can be attributed
to direct Compton processes with $x_{\gamma}=1$ with the
remaining 12~\% due to resolved events and 13~\% due to radiative events
where one of the outgoing quarks in a dijet event radiates
a photon. More data are needed to obtain a more quantitative
constraint for the quark distribution in the photon.

\section{\boldmath $D^{*\pm}$
production in  $\gg$ and $\gamma$p interactions
\unboldmath}
In a similar manner to jet production, open charm production in $\gg$ and
$\gamma$p collisions 
can also be used to constrain the parton content of the photon.
The charm production cross-sections have been calculated
in NLO for $\gg$~\cite{bib-drees,bib-cacc1} 
and $\gamma$p interactions~\cite{bib-frix,bib-ckniehl,bib-cacc2}. 
The NLO calculations are either done in the so-called
massive~\cite{bib-drees,bib-frix} or in the so-called
massless~\cite{bib-cacc1,bib-ckniehl,bib-cacc2} scheme.
\begin{figure}[htbp]
\begin{picture}(15.0,6.5)
\put(0,3.25){\begin{tabular}{cc}
\epsfig{file=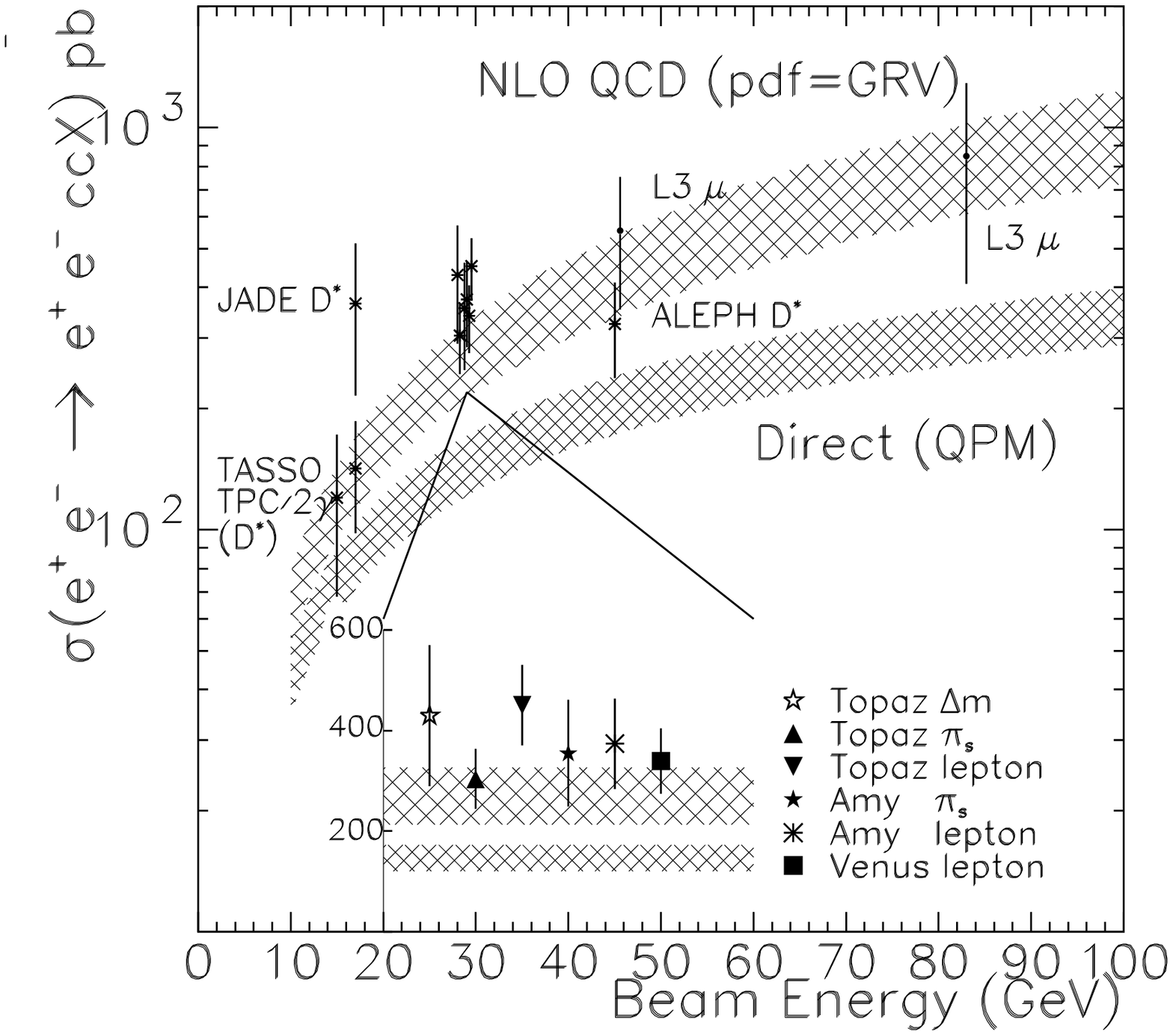,width=0.49\textwidth,
height=5.5cm}
 &
\epsfig{file=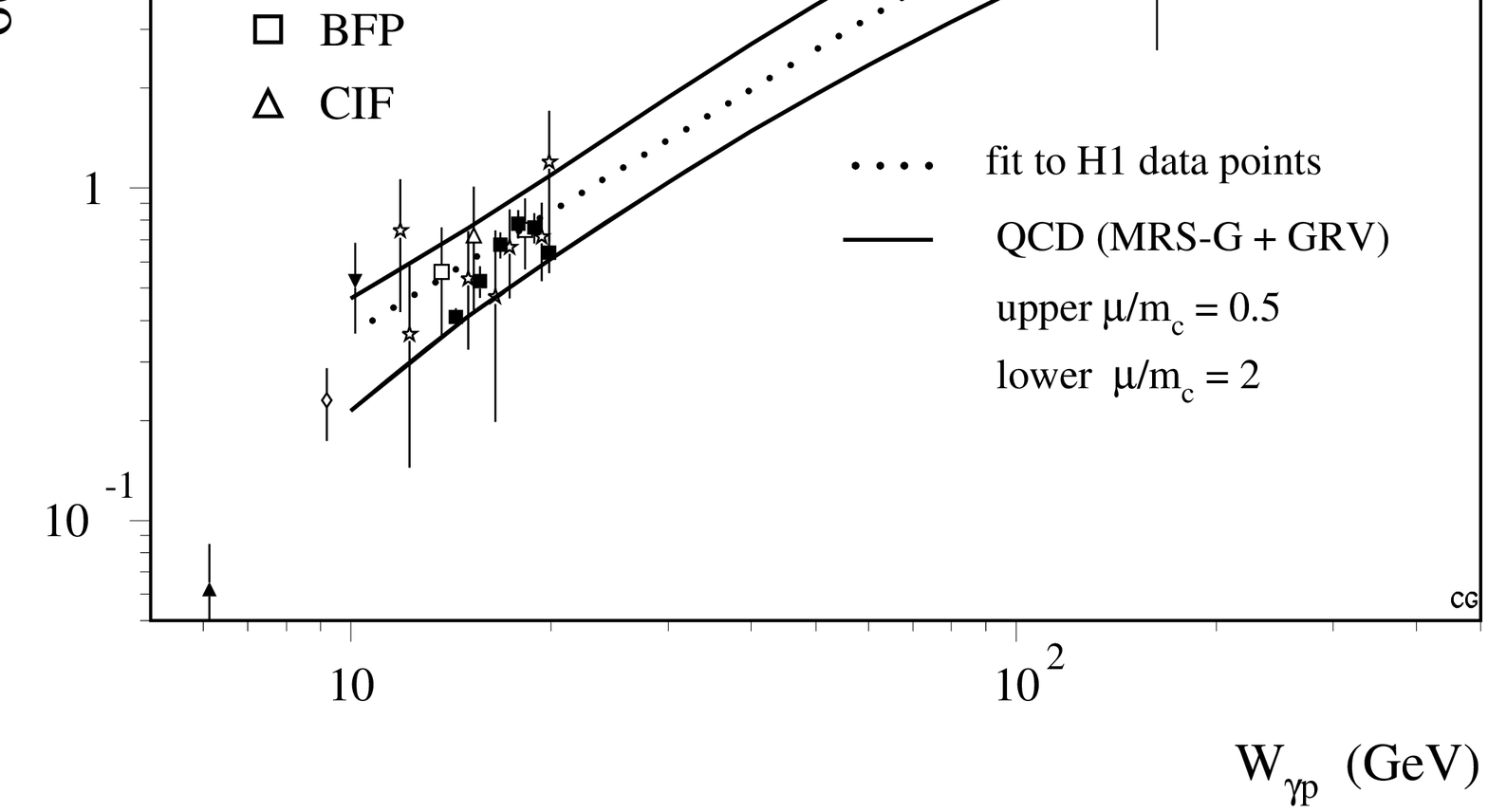,width=0.43\textwidth,
height=3.2cm} 
\end{tabular}}
\put(4.2,0.35){(a)}
\put(12.0,0.35){(b)}
\end{picture}
\caption{\label{fig-ctot} 
a) Cross-section for the process $\ee\rightarrow\ee\ccbar X$ as a function of the 
electron beam energy;
b) cross-section for the process $\gamma\mbox{p}\rightarrow\ccbar X$ 
as a function of the $\gamma$p centre-of-mass energy $W_{\rm \gamma p}$).
}
\end{figure}
 
In the massive scheme the mass $m_{\rm c}$ of the charm quark sets
the scale for the perturbative QCD calculation. The cross-section
is factorized into the matrix elements for the production
of heavy quarks and the parton densities for light quarks (q) and
gluons. This `massive' approach is expected to be valid
if the transverse momenta $p_{\rm T}$ of the charm quarks are of the
same order, $p_{\rm T} \approx m_{\rm c}$.
At LEP energies only the direct process $\gg\rightarrow\ccbar$ and the
single-resolved process gq$\rightarrow\ccbar$ are important.
The relevant processes at HERA are gq$\rightarrow\ccbar$ 
and gg$\rightarrow\ccbar$. The number of gluon initiated processes
depends very much on the parametrisations of the parton densities used.
Again, only the sum of the event classes called resolved 
\begin{wrapfigure}[20]{r}{2.0in}
\epsfig{file=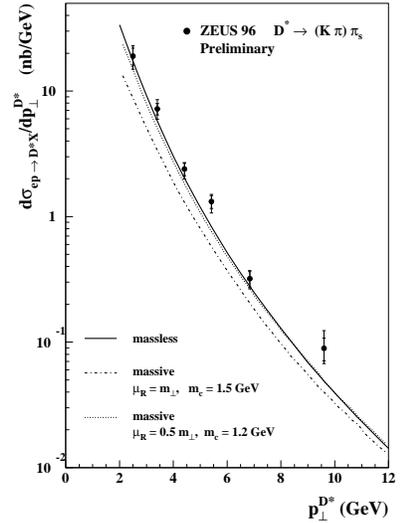,width=2.0in}
\caption{\label{fig-charmpt}
Differential cross-section ${\rm d}\sigma/{\rm d}p_{\rm T}^{D^*}$
in the kinematic region $Q^2<4$~GeV$^2$, $115<W<280$~GeV and
$-1.5<\eta^{D*}<1.0$}  
\end{wrapfigure}
and direct is well defined in NLO.
In the `massless' scheme charm is considered as one of
the active flavours in the parton distributions like u,d,s.
This scheme is expected to be valid for $p_{\rm T} >> m_{\rm c}$.

The cleanest method to tag open charm is the reconstruction of 
$D^{*+}\rightarrow D^0\pi^+$ decays. Due to the small branching ratios
of the $D^0$ into charged pions and kaons, this method is statistics limited. 
ALEPH has measured the charm cross-section 
$\sigma(\ee\rightarrow\ee\ccbar X)$
using $33\pm8$ $D^{*\pm}$ mesons reconstructed in their LEP1 
data~\cite{bib-alcharm} with $P_{\rm T}^{D^*}>2$~GeV.
L3 has measured the charm cross-section in $\gg$ interactions
at LEP1 and LEP2 by tagging muons 
from semi-leptonic charm decays
in the momentum range $2<p_{\mu}<7$~GeV$/c$ at LEP1 and
$2<p_{\mu}<15$~GeV/$c$ at LEP2~\cite{bib-l3charm}. The efficiency to tag muons
is less than $10^{-3}$ leading to large systematic and
statistical uncertainties. 

The cross-section for the process $\ee\rightarrow\ee\mbox{c}
\overline{\mbox{c}}$ as a function of the beam energy is shown
in Fig.~\ref{fig-ctot}a. The experimental results for
various charm tagging methods used by pre-LEP
experiments have been extrapolated to obtain a total charm
cross-section~\cite{bib-lep2}. 
The upper band shows the full NLO charm cross-section
calculated in the massive scheme by Drees et al~\cite{bib-drees} and
the lower band the contribution from the Born term direct process
(Quark Parton Model, QPM). The upper edge of the band
is obtained by setting $m_{\rm c}=1.3$~GeV with a  scale $\mu=m_{\rm c}$ 
and the lower edge by setting $m_{\rm c}=1.7$~GeV with $\mu=\sqrt{2}
m_{\rm c}$.
The data points obtained from the TRISTAN and JADE measurements cluster
around the higher edge of the the massive NLO calculation 
which uses the GRV parametrisation. 

H1~\cite{bib-h1charm} and ZEUS~\cite{bib-zeuscharm1,bib-zeuscharm2} have
measured $D^{*\pm}$ production cross-sections in $\gamma$p interactions.
Both experiments have derived a cross-section for
the process $\gamma\mbox{p}\rightarrow\ccbar X$ as a function
of the $\gamma$p centre-of-mass energy 
$W_{\rm \gamma p}$~\cite{bib-h1charm,bib-zeuscharm2}.
In Fig.~\ref{fig-ctot}b this cross-section is compared to
the results of lower energy experiments and to a massive NLO
calculation~\cite{bib-frix} using the NLO GRV parametrisation.
Within the large errors the massive NLO calculations using the
GRV parton distributions are in good agreement with the total 
charm production cross sections measured at LEP and HERA. The
cross-section depends very much on the gluon distribution
in the photon, e.g. the LAC1 parametrisation gives a much
larger cross-section.

The extrapolation of the $D^{*\pm}$ production cross-section
to a total $\ccbar$ cross-section has large theoretical
and experimental uncertainties. These are avoided if
differential distributions are measured.
ZEUS~\cite{bib-zeuscharm1} has measured the 
$p_{\rm T}^{D^*}$ distribution (Fig.~\ref{fig-charmpt}) and compared it
to a massive NLO calculation~\cite{bib-frix} using the
GRV parametrisation for the photon and to a massless
calculation~\cite{bib-ckniehl}. The normalisation of the massless calculation
which uses similar parameters as in the case of the massive calculation 
with a charm mass $m_{\rm c}=1.5$~GeV is in better agreement with the data.

\section{Total cross sections}
\label{sec-total}
The total cross-sections $\sigma$ for hadron-hadron and $\gamma$p 
collisions are well described by a Regge parametrisation of the form
$\sigma=X s^{\epsilon}+Y s^{-\eta}$,
where $\sqrt{s}$ is the centre-of-mass energy of the hadron-hadron
or $\gamma$p interaction. The first term in the equation
is due to Pomeron exchange and the second term
is due to Reggeon exchange~\cite{bib-gallo}. Assuming factorisation of the 
Pomeron term $X$, the total hadronic $\gg$ cross-section $\sigma_{\gg}$
can be related
to the pp (or $\ppbar$) and $\gamma$p total cross-sections at 
high centre-of-mass energies $W=\sqrt{s}_{\gg}$ where the Pomeron trajectory
should dominate:
\begin{equation}
\sigma_{\gg}
=\frac{\sigma_{\gamma{\rm p}}^2}{\sigma_{\rm pp }}.
\label{eq-tot2}
\end{equation}
A slow rise of the total cross-sections with energy is predicted,
corresponding to $\epsilon\approx0.08$.
This rise can also be
attributed to an increasing cross-section for parton interactions
leading to mini-jets in the final state~\cite{bib-minijet}.

Before LEP $\sigma_{\gg}(W)$ has been measured
by PLUTO~\cite{bib-pluto}, TPC/2$\gamma$~\cite{bib-tpc} and
MD1~\cite{bib-md1}. These experiments have measured
at $\gg$ centre-of-mass energies $W$ below 10~GeV before the onset 
of the high energy rise of the total cross-section.
Using LEP data taken at $\sqee=130-161$~GeV
L3~\cite{bib-l3tot} has demonstrated that
$\sigma_{\gg}(W)$ is consistent with the universal Regge behaviour of
total cross-sections in the range $5\le W \le 75$~GeV 
\begin{figure}[htbp]
\begin{picture}(15.0,6.0)
\put(0,3.0){\begin{tabular}{cc}
\epsfig{file=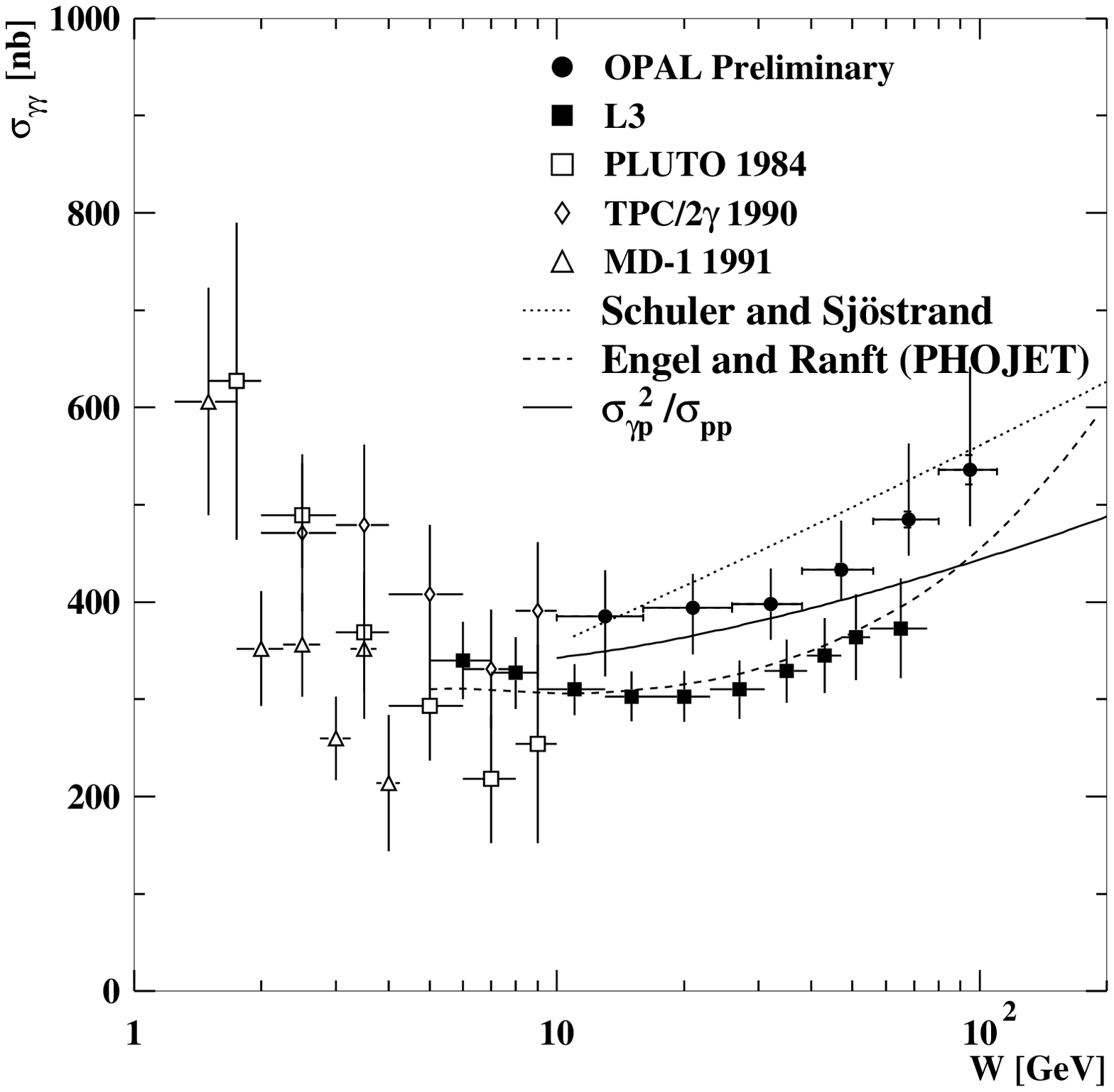,width=0.46\textwidth,height=6.0cm}
 &
\epsfig{file=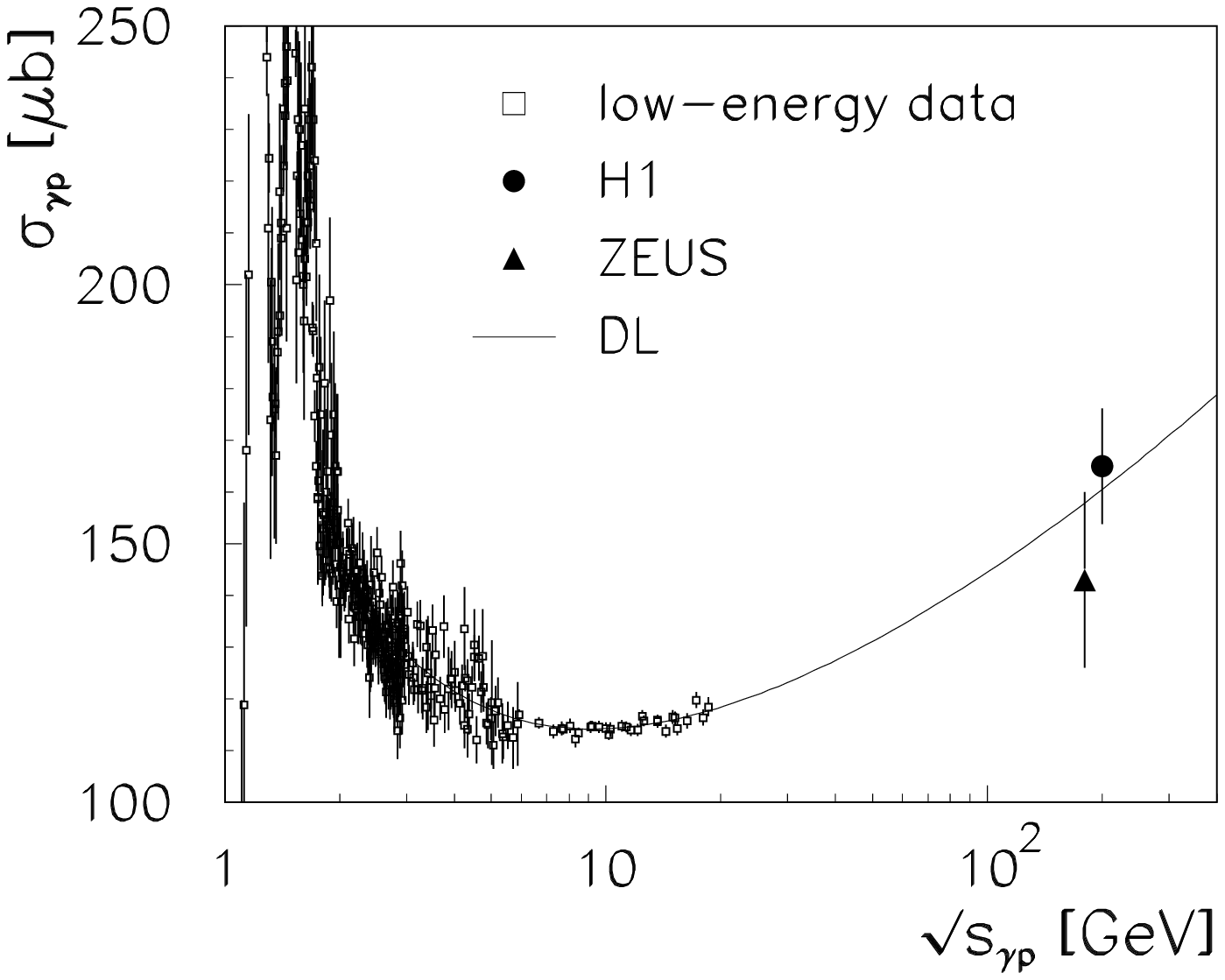,width=0.46\textwidth,height=6.0cm} 
\end{tabular}}
\put(1.5,5.35){(a)}
\put(9.8,5.35){(b)}
\end{picture}
\caption{\label{fig-stot} 
Total cross-section of the process (a) $\gg\rightarrow\mbox{hadrons}$
as a function of $W=W_{\gg}=\sqrt{s}_{\gg}$
and of the process (b) $\gamma\mbox{p}\rightarrow\mbox{hadrons}$
as a function of $W_{\gamma p}=\sqrt{s}_{\gamma\rm p}$.}
\end{figure}
which was also observed in $\gamma$p scattering at HERA.
The L3 measurement is shown in Fig.~\ref{fig-stot}a
together with an OPAL measurement~\cite{bib-frank} in the range
$10<W<110$~GeV using data taken at $\sqee=161-172$~GeV.
The observed energy dependence of the cross-section is similar,
but the values for $\sigmagg$ are about 20~\% higher. 
The errors are strongly correlated between the $W$ bins
in both experiments. Furthermore, L3 has used the Monte Carlo
generator PHOJET~\cite{bib-phojet} for the unfolding, whereas OPAL has 
averaged the unfolding results of PHOJET and PYTHIA.
The unfolded cross-section using PHOJET
is about 5~\% lower than the central value.
In both experiments the cross-sections obtained
using PHOJET are lower than the cross-section obtained
with PYTHIA. The origin of the remaining discrepancy is
not yet understood.

Based on the Donnachie-Landshoff (DL) model~\cite{bib-DL}, 
the assumption of a universal high energy behaviour of 
$\gg$, $\gamma$p  and pp cross-sections is tested.
The parameters $X$ and $Y$ are fitted to the
total $\gg$, $\gamma$p and pp cross-sections 
in order to predict $\sigmagg$ via Eq.~\ref{eq-tot2}.
This is done assuming that the cross-sections can
be related at $\sqrt{s}_{\gg}=\sqrt{s}_{\rm \gamma p}=\sqrt{s}_{\rm pp }$.
The process dependent fit values for $X$ and $Y$ are
taken from Ref.~73 together with the values of the universal
parameters $\epsilon = 0.0790 \pm 0.0011$ and 
$\eta = 0.4678 \pm 0.0059$.
This simple ansatz gives a reasonable
description of the total $\gg$ cross-section $\sigmagg$.
Schuler and Sj\"ostrand~\cite{bib-GSTSZP73} give a 
total cross-section for the sum of all possible event
classes in their model of $\gg$ scattering where the photon
has a direct, an anomalous and a VDM component.
They consider the spread between this prediction and
the simple factorisation ansatz as conservative estimate
of the theoretical band of uncertainty.
The prediction of Engel and Ranft~\cite{bib-phojet} is also plotted
which is implemented in PHOJET. It is in good agreement with
the L3 measurement and significantly lower than the OPAL
measurement. 
The steeper rise predicted by Engel and Ranft is in agreement
with both measurements.

\section{Photon fragmentation function}
Closely related to the parton distributions of the photon
are the fragmentation functions $D_{\rm q,g}^{\gamma}(z,M^2)$
of quarks and gluons into photons with
$z$ being the fractional momentum carried by the photon.
The photon fragmentation functions $D_{\rm q,g}^{\gamma}(z,M^2)$
are measured in $\ee$ annihilation at LEP. The time-like scale
$M^2$ is therefore given by the $\ee$ centre-of-mass energy.

As in the case of the photon structure function, the photon 
fragmentation function is fully calculable in perturbative
QCD for asymptotically large $M^2$ due to the pointlike coupling of the 
photon to $\qqbar$ pairs. At experimentally accessible
values of $M^2$ the non-perturbative contributions
are still large and have to be taken into account.
Bourhis et~al.~\cite{bib-bourhis} have calculated new
fragmentation functions with
a full treatment of the Beyond Leading Logarithm (BLL) corrections
to the perturbative part and with a VDM input for the non-perturbative
part.

\begin{wrapfigure}[18]{r}{0.415\textwidth}
\epsfig{file=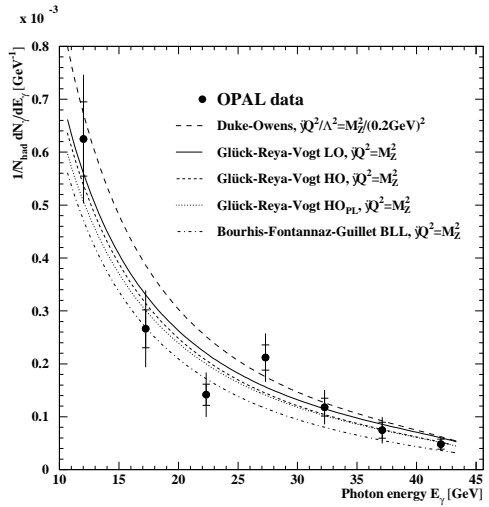,width=0.415\textwidth} 
\caption{\label{fig-frag}
Distribution of the photon energy $E_{\gamma}$ in hadronic
$\Zzero$ decays.}
\end{wrapfigure}
A first measurement of the quark-to-photon fragmentation function 
has been published by ALEPH using isolated photons as well as photons
which were reconstructed inside hadronic jets~\cite{bib-alfrag}.
It was pointed out by the authors of Ref.~75
that these measurements cannot be compared directly to
the fully inclusive calculations of the photon fragmentation
functions due to phase space restrictions imposed by
the jet algorithm used. OPAL~\cite{bib-opalfrag} has therefore measured
the fully inclusive energy spectrum for 
photons with energy $E_{\gamma}>10$~GeV in hadronic $\Zzero$ decays. The huge
background from photons due to hadron decays (e.g.~$\pi^0\rightarrow
\gg$) is subtracted using a fitting method. The data
shown in Fig.~\ref{fig-frag} are in agreement with
the models by Bourhis et al~\cite{bib-bourhis}, 
Duke and Owens~\cite{bib-dofrag} and by Gl\"uck et~al.~\cite{bib-grvfrag}. 
The differences between the GRV LO calculation and the higher order (HO) 
calculation with and without the non-perturbative corrections ($\rm HO_{PL}$)
is much smaller than the experimental errors.

\section{Conclusions}
Interactions of photons via quantum fluctuations can be described 
using a structure functions formalism. 
The hadronic structure function $F_2^{\gamma}(x,Q^2)$ of the photon is
measured in e$\gamma$ scattering at LEP in the range $x>10^{-3}$ and 
$1<Q^2<10^3$~GeV$^2$.
The logarithmic rise of $F_2^{\gamma}$ with $Q^2$ for
medium $x$ and large $Q^2$ is observed as predicted by perturbative QCD.
At low $x$ LEP will be able to study the region where the onset of the rise of
$F_2^{\gamma}$ is expected from the HERA data on the
proton structure function. 

The parton content of the photon is also measured
in $\gg$ and $\gamma$p interactions at LEP and HERA.
Jet production and high $p_{\rm T}$ hadron production 
are especially sensitive to the gluon distribution in the photon.

The GRV, SaS, GS and LAC parton distributions of the photons
are currently the most widely used parametrisations.
Parametrisations with a large gluon distribution like LAC1 are disfavoured
by the data. The other distributions are consistent
with most of the available measurements.

H1 has presented new results on the structure of virtual photons
from the $Q^2$ dependence of jet production. More information
about the interactions of photons at low and medium $Q^2$
are to be expected in the near future from the measurement
of double-tagged $\gamma^*\gamma^*$ events at LEP.

First measurements by L3 and OPAL of the energy dependence of the total
$\gg$ cross-section for hadron production show the rise characteristic
of hadronic interactions.

\section*{Acknowledgements}
I want to thank the HERA and the LEP collaborations for providing all the 
interesting results. My apologies to the CLEO and to the L3 collaboration
for not covering their results on form factors and glueball searches.
I am also very grateful to Jon Butterworth, John Dainton, Albert De Roeck,
Martin Erdmann, Alex Finch, Maria Kienzle, Michael Klasen, Aharon Levy and
Richard Nisius for their support and advice.

\end{document}